\documentclass[12pt,a4paper]{JHEP3}
\usepackage{amsmath,amssymb,bm}
\usepackage{latexsym}
\usepackage{graphicx}

\def\eqref#1{Eq.~(\ref{#1})}

\def\Eq#1{\begin{equation} #1 \end{equation}}
\def\Eqr#1{\begin{eqnarray} #1 \end{eqnarray}}
\def\Eqrsubl#1#2{\begin{subequations}\label{#1}\Eqr{#2}\end{subequations}}

\newcommand{\nn}{\nonumber}

\def\Usp{{\rm U}}

\def\bMsp{\bar{{\rm M}}}

\def\Zsp{{\rm Z}}

\def\X5sp{{\rm X}_5}
\def\Y3sp{{\rm Y}_3}
\def\Z3sp{{\rm Z}_3}

\def\lap{{\triangle}}
\def\e{{\rm e}}

\newcommand{\bea}{\begin{eqnarray}}
\newcommand{\eea}{\end{eqnarray}}
\newcommand{\beq}{\begin{equation}}
\newcommand{\eeq}{\end{equation}}

\title{Moduli Stabilization in a de Sitter Compactification Model}

\author{
Antonino Flachi\\
Multidisciplinary Center for Astrophysics, 
Instituto Superior Tecnico,\\
~~Lisbon, 1049-001, Portugal.\\
~~~E-mail: \email{antonino.flachi$``$at$"$ist.utl.pt}}
\author{
Masato Minamitsuji\\
Multidisciplinary Center for Astrophysics, 
Instituto Superior Tecnico,\\
~~Lisbon, 1049-001, Portugal.\\
~~~E-mail: \email{masato.minamitsuji$``$at$"$ist.utl.pt}}
\author{
Kunihito Uzawa\\
Department of Physics,
School of Science and Technology,\\
~~Kwansei Gakuin University, Sanda, Hyogo 669-1337, Japan.\\
~~~E-mail: \email{uzawa$``$at$"$yukawa.kyoto-u.ac.jp}}

\abstract{%
We discuss the moduli stabilization in a de Sitter compactification model
obtained coupling $D$-dimensional gravity to scalar and gauge 
fields. 
This class of models is characterized by two  
moduli: one related to the volume of the internal space, the other to the 
warp factor. 
While the volume modulus can be fixed by appropriately tuning the gauge 
field strength, 
curvature of the internal space, and cosmological constant, the same 
mechanism does not work for the warp modulus.
In this paper we discuss a stabilization mechanism based on quantum effects and show that both moduli can be efficiently stabilized.
}

\begin{document}


\section{Introduction}
 \label{sec:introduction}
The recent discovery of dark energy demands a mechanism 
setting the cosmological constant to a value that is nonzero 
but hierarchically small compared to the Planck scale.  
At the same time, a lot of recent observational data,
in particular those of the cosmic microwave background,
support the basic predictions of inflationary scenarios. 
A theoretical framework that may be able to provide a consistent description 
of the universe, undergoing inflation at early times and dominated 
by dark energy at the present day, is offered by string theory.
In general, string theory requires the presence of extra dimensions  
that have to be stabilized at some appropriate scale
to obtain a viable cosmological model. 
The lack of such a mechanism is often called 
the moduli stabilization problem. 
The stabilization of moduli is 
deeply connected with
the realization of 
above accelerating phases
in the cosmic history,
and plays an important role in the construction of higher-dimensional
cosmological models 
(See e.g., \cite{fr,gw,cghw,gl}). 

While the problem of moduli stabilization has been discussed at length 
for the case of Kaluza-Klein compactifications, its analysis for warped 
compactifications remains much less extensive \cite{Kodama:2005fz, 
Kodama:2005cz, Kodama:2006ay}. In fact, the case of warped compactifications 
is an interesting set-up to consider since, due to the warping and external 
directions, multiple scalar moduli may appear and mix in a non-trivial way, 
affecting, in principle, each other's dynamics.  
The present paper aims at discussing an example of this sort.

Specifically, we will consider gravity coupled to a scalar dilaton and a form field strength propagating on the background of a higher-dimensional 
warped geometry of the form 
\bea
ds^2 &=& \e^{2 A(y)} \left[q_{\mu\nu}(\Usp) du^{\mu}du^{\nu} 
+ dy^2 + 
{\gamma}_{ab}(\Zsp) dz^adz^b \right],
\label{I:metric:Eq}
\eea
where $ds^2(\Usp)=q_{\mu\nu}(\Usp) du^{\mu}du^{\nu}$ and 
$ds^2(\Zsp)={\gamma}_{ab}(\Zsp) dz^adz^b$ represent, respectively, the line elements of two maximally symmetric, non-singular manifolds U and Z, and $y$ is the direction of warping. The coordinates $u^\mu$ and $z^a$ parametrize, respectively, the manifolds $\Usp$ and $\Zsp$.
Z is assumed to be compact. The dimensionalities of U and Z are, respectively, $n$ and $D-n-1$. Solutions of the above type 
have been discussed in Ref.~\cite{Minamitsuji:2011gp} and will be briefly recalled here for the convenience of the reader.

The class of models that will be considered in this work is described, 
in the Einstein frame, by the following action
\Eq{
S=\frac{1}{2\kappa^2}\int \Big[\left\{R
-2\e^{-\alpha\phi/(p-1)}\Lambda\right\}\ast{\bf 1}
-\frac{1}{2}d\phi\wedge\ast d\phi
-\frac{1}{2\cdot p!}\e^{\alpha\phi}F \wedge\ast F\Big],
\label{fs:action:Eq}
}
where $\kappa^2$ is the $D$-dimensional gravitational constant, 
$\ast$ is the Hodge operator in $D$-dimensions, 
$\phi$ is a scalar field, 
$F$ is a $p$-form field strength, and $\Lambda$ and $\alpha$ are constants.

The equations of motion follow directly from the above action. The $p$-form field strength is taken to be proportional to the volume form of $\Zsp$,
$$
\Omega(\Zsp)=\sqrt{\gamma}dz^1\wedge\cdots\wedge dz^p\,,
$$
that is
\begin{equation}
F = f\,\Omega(\Zsp)\,,~~~~p=D-n-1\,,
    \label{fs:strength:Eq}
\end{equation}
with $f$ constant and $\gamma$ denoting the determinant of the metric $\gamma_{ab}$. The choice (\ref{fs:strength:Eq}) gua\-ran\-tees that both the Bianchi identities and the equation of motion for the gauge field are automatically satisfied.

The ansatz for the scalar field $\phi$ is
\begin{equation}
\phi = \frac{2}{\alpha}(p-1)A(y)\,,
    \label{fs:scalar:Eq}
\end{equation}
leading to the following equation:
\Eqr{
A''+(D-2)\left(A'\right)^2
-\frac{\alpha^2 \hat{\Lambda}}{p-1}
=0\,,
   \label{fs:scalar3:Eq}
}
where $\hat{\Lambda}=-\frac{\Lambda}{p-1}+\frac{f^2}{4}$ and $'$ denotes the ordinary derivative with respect to the coordinate $y$. 
Finally, using the metric ansatz (\ref{I:metric:Eq}), Einstein equations can be expressed as 
\Eqrsubl{fs:Einstein2 eq:Eq}{
&&R_{\mu\nu}(\Usp)-\beta\hat{\Lambda} q_{\mu\nu}(\Usp)=0\,,
   \label{fs:Einstein-mn2:Eq}\\
&&A''+\frac{2(p-1)^2}{(D-1)\alpha^2}\left(A'\right)^2
-2 
\frac{(p-1)}{(D-1)}
\hat{\Lambda}
=0\,,
   \label{fs:Einstein-yy2:Eq}\\
&&R_{ab}(\Zsp)-\left(\beta\hat{\Lambda}+\frac{f^2}{2}\right)\gamma_{ab}(\Zsp)=0\,,
   \label{fs:Einstein-ab2:Eq}
}
where $R_{\mu\nu}(\Usp)$ and $R_{ab}(\Zsp)$ are the Ricci 
tensors of the metrics $q_{\mu\nu}$ and $\gamma_{ab}$, respectively, 
and the constant $\beta$ is defined by
\Eq{
\beta=\frac{\alpha^2}{p-1}-\frac{2(p-1)}{D-2}\,.
   \label{fs:beta:Eq}
}
Off-diagonal components of the Einstein equations are automatically satisfied by our ansatz. Eqs.(\ref{fs:scalar3:Eq}) and (\ref{fs:Einstein-yy2:Eq}) can be simultaneously solved 
as
\Eq{
A(y)=\ell\left(y-y_0\right)\,,
\label{abg}
}
with $v_0$ constant and $\ell$ given by
\Eq{
\ell=\pm\,\alpha\,\sqrt{\frac{\hat{\Lambda}}{(p-1)(D-2)}}\,.
}
Notice that the above solution is only compatible 
with the condition $\alpha^2 \neq 2(p-1)^2/(D-2)$.
Choosing $\alpha$ such that $\beta>0$ 
ensures that both $\Usp$ and $\Zsp$ are positively curved,
 as it is clear from an inspection of \eqref{fs:Einstein-ab2:Eq}. 
This corresponds to taking
\Eq{
\alpha>\sqrt{\frac{2}{D-2}}\,(p-1)\,,~~~~~~
\alpha<-\sqrt{\frac{2}{D-2}}\,(p-1)\,.
}
In this case, the field equations lead to the following solution 
for the $D$-dimensional metric 
\Eq{
ds^2=\e^{2\ell(y-y_0)}\left[
q_{\mu\nu}^{({\rm dS})} du^\mu du^\nu
+dy^2+\gamma_{ab}(\Zsp)dz^adz^b\right]\,,
  \label{fs:D-metric2:Eq}
}
where $q_{\mu\nu}^{({\rm dS})}$ represents the metric of de Sitter space with expansion rate $H^2 ={\beta \hat{\Lambda}\over n-1}$ as it follows from Eq.~(\ref{fs:Einstein-mn2:Eq}). 



Details will be given later, here we 
simply mention that starting from the above background solution, an effective theory can be directly derived 
by compactifying the $\Zsp$ space. In the simplest construction, 
the effective theory contains
two unstabilized moduli: one related to the volume of the internal 
space $\Zsp$, and another to the warp factor.
As we will see, simultaneous 
stabilization of both moduli may not be achieved by means of the same 
mechanism. For instance, appropriately tuning the gauge field strength, 
curvature of the spherical internal space, and cosmological constant may 
help to achieve stabilization of the volume modulus but not of 
the warp factor \cite{Minamitsuji:2011gp, Minamitsuji:2011gn}.

A different stabilization mechanism can be constructed by using quantum 
effects. In this case, if the volume modulus is stabilized due to the 
presence of a gauge field strength, its quantum fluctuation as well as those 
of the modulus associated to the warp factor may both contribute to 
stabilize or destabilize the background geometry. 
After flux stabilization, the volume modulus fluctuates around the 
minima of the effective potential and its contribution can be computed in 
a straightforward manner. On the other hand, at tree level, the dynamics of 
the warp modulus is controlled by a runaway type of potential, 
and its quantum fluctuations should be analyzed with care. 
In the following, we will adopt a self-consistent approach
and require that any acceptable minima of the effective 
potential must occur where the potential is sufficiently 
flat.

An important point to remark is related to the value of the scalar 
potential after stabilization. 
In principle, once quantum corrections are 
included, the scalar potential of the system may occur at a 
positive, vanishing 
or negative value, resulting in a de Sitter, Minkowski or anti de Sitter 
geometry. In this case, we may expect that additional corrections to the 
potential, for example due to finite temperature effects, may produce a 
further shift up-lifting its minima from anti-de Sitter to Minkowski or 
de Sitter, or, at very high temperature, pushing the system into an 
unstabilized phase. 

The paper is organized as follows. In Sec.~\ref{sec:flux}, we will present 
the model in detail and construct the effective theory tuning the field 
strength to achieve stabilization of the volume modulus at the classical 
level. The main part of the paper is devoted to discuss how quantum effects 
from moduli contribute to the effective potential at one-loop. 
We will adopt the background field method and path integrals to perform 
the computation and use a zeta function regularization. 
Specifically, Sec.~\ref{sec:mass} deals with the contribution from the warp modulus to see whether its quantum fluctuations may provide any 
stabilization. 
In fact, due to the runaway behavior of the potential, as we have already 
mentioned, the minima (if any) generated by quantum fluctuations must be in 
a region where perturbation theory can be trusted. This self-consistency 
requirement is then verified {\it a posteriori}. 
After presenting the machinery we will perform the 
computation using an approach based on contour integral techniques similar 
to that described in Refs.~\cite{Uzawa:2003ji, Uzawa:2003qh, 
Kikkawa:1984rx, Kikkawa:1984qc} (Related work is that of 
Refs.~\cite{Flachi:2004hc, Flachi:2003bb, Flachi:2003bn, Eliz, Milton}). 
This method is valid over the whole parameter space and serves as a general 
way to compute the one-loop effective potential. 
In a restricted range of the parameter space a slightly simplified 
approach based on the Schwinger-De Witt approximation can be adopted. 
This method uses directly the 
small-$t$ heat-kernel asymptotics and it applies only in a small region 
of the parameter space. Details of this second approach will be given in 
Appendix~\ref{SDW} where the validity of the Schwinger-De Witt approximation 
will also be discussed. Results using both method are consistent when 
applied to the same region of the parameter space. 
(In Appendix \ref{fta} we will show how finite temperature corrections may 
produce transitions between different minima uplifting the vacuum. 
These effects are studied by means of the standard Matsubara formalism.) 
We will show that quantum corrections from the warp modulus 
can provide stabilization and lead to a de Sitter, Minkowski or anti 
de Sitter minimum. Unfortunately, the region of the parameter space 
for the moduli-stabilization
consistent with the semi-classical approximation 
is only marginal. 
In Sec.~\ref{sec:mod} we add the contribution to the potential from   
from quantum fluctuations of the volume modulus again using an approach 
based on contour integrals and show that this may provide an efficient 
framework for stabilization.
This seems rather natural, since after flux compactification the 
size of the internal space generically becomes of order of the Planck length. 
In this case, it is not possible to ignore quantum fluctuations of the
volume modulus, even though the volume is already stabilized. These 
contributions to the one-loop effective potential may stabilize the 
warped direction and naturally realize a de Sitter, Minkowski or anti de 
Sitter minimum depending on the values of the parameters and of the 
renormalization scale. Our conclusions close the paper.

\section{The effective theory with field strengths}
\label{sec:flux}

The effective theory will be constructed in this section by promoting the warp factor and the size of the external manifold $\Zsp$ to scalar degrees of freedom. To do so, we express the metric (\ref{I:metric:Eq}) in the following way:
\Eq{
ds^2=\e^{2\bar{A}(u_\mu, y)}\left[q_{\mu\nu}(\Usp) du^{\mu}du^{\nu} + dy^2
+\e^{2\bar{\psi}(u_\mu, y)}\gamma_{ab}(\Zsp)dz^adz^b\right]\,.
  \label{fs:metric:Eqss}
}
The background solution discussed in the previous section corresponds to the above metric once $\bar{A}=A(y)$ as given in (\ref{abg}) and $\bar{\psi}= constant$.

\begin{figure*}[ht]
\unitlength=1.1mm
\begin{center}
\begin{picture}(120,37)
   \put(-5.,20){\rotatebox{90}{$W(\bar \psi)$}}
   \put(61,20){\rotatebox{90}{$W(\bar \psi)$}}
   \put(29,-3){$\bar \psi$}
   \put(97,-3){$\bar \psi$}
   \put(105,30.5){\tiny{$f=0$}}
   \put(105,26){\tiny{$f=98$}}
   \put(105,21){\tiny{$f=118$}}
   \put(105,16.5){\tiny{$f=180$}}
   \put(105,11){\tiny{$\lambda=0.5$, $\Lambda=0.2$}}
   \put(36,13.2){\tiny{$\lambda=0.01$}}
   \put(36,10.5){\tiny{$\lambda=0.06$}}
   \put(36,8){\tiny{$\lambda=0.11$}}
   \put(36,5.2){\tiny{$\lambda=0.16$}}
   \put(38,3){\tiny{$f=0$, $\Lambda=1$}}
  \includegraphics[height=3.9cm]{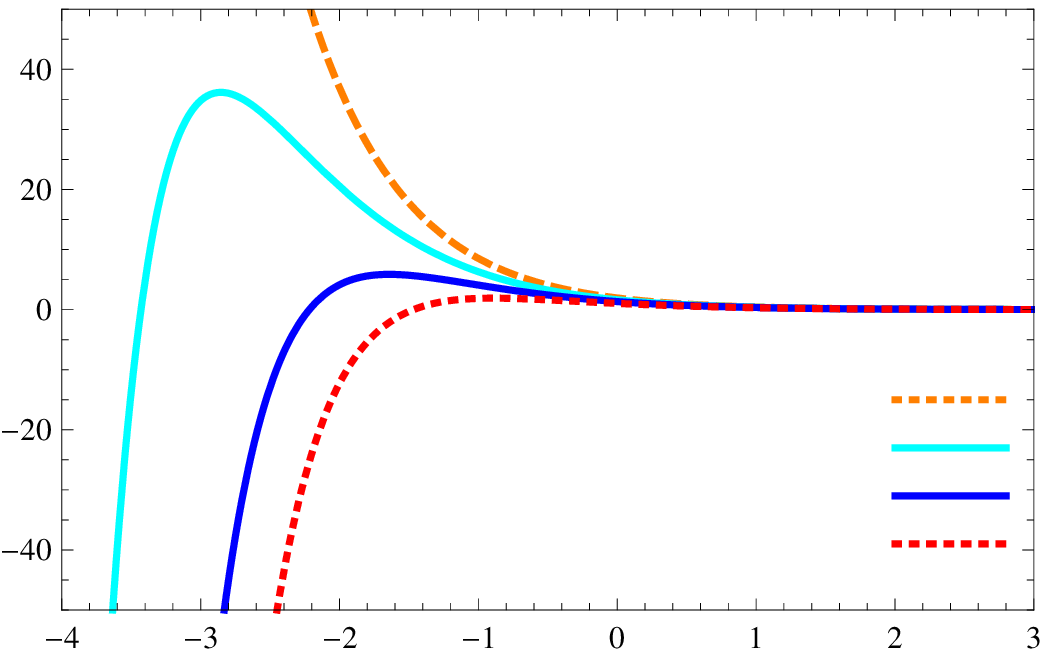}
  \hskip 1cm
  \includegraphics[height=3.9cm]{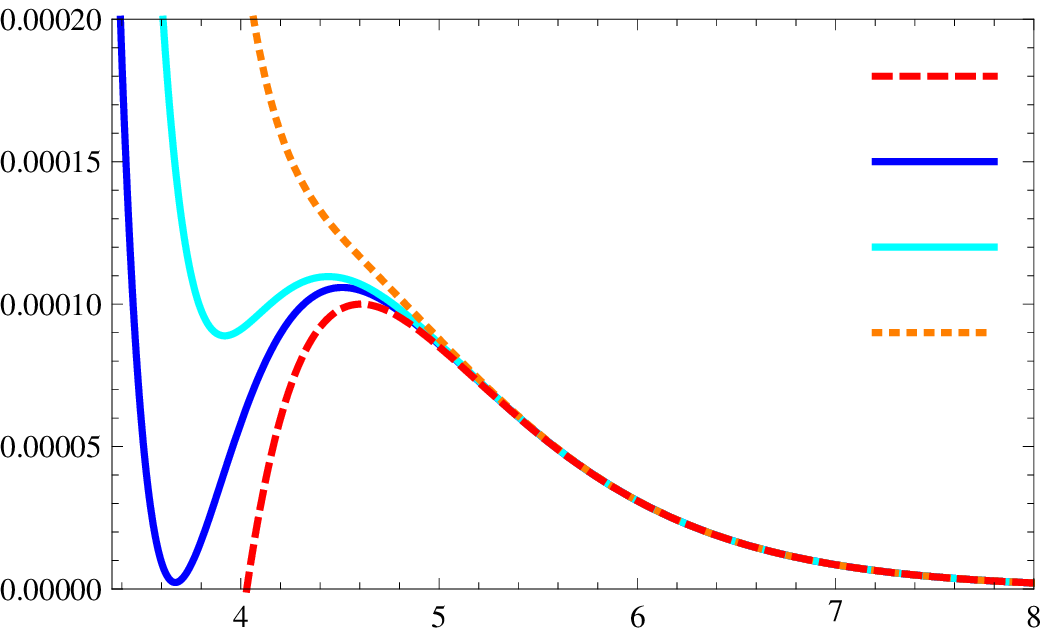}
\end{picture}
  \caption{
The figure illustrates the behavior of the potential $W(\bar\psi)$.
The left panel displays the potential for vanishing 
gauge field, $f=0$, and for several values of the constant $\lambda$ 
(with $\Lambda$ normalized to unity). The right panel shows the potential 
for several choices of $f$ (with the values set as indicated in the figure). 
Dimensionality parameters are chosen as follows: $D=10$, $p=6$ and $n=3$.
}
  \label{vpsi}
\end{center}
\end{figure*}

Using (\ref{fs:strength:Eq}), (\ref{fs:scalar:Eq}) and (\ref{fs:metric:Eqss}) 
in the $(n+1)$-dimensional action (\ref{fs:action:Eq}), after 
using the equation of motion for the background solution, we get  
\Eqr{
S&=&\frac{1}{2\tilde{\kappa}^2} \int_{\bMsp}
 \left[\left\{R(\bMsp)-V\left(\bar{A},~\bar{\psi}\right)\right\}
 \ast_{\bMsp}{\bf 1}_{\bMsp}-\frac{1}{2}d\bar{A}\wedge\ast_{\bMsp}d\bar{A}
\right. \nn\\ 
&&\left.
-\frac{1}{2}\frac{c_2}{\sqrt{c_1c_3}}d\bar{A}\wedge\ast_{\bMsp}d\bar{\psi}
 -\frac{1}{2}d\bar{\psi}\wedge\ast_{\bMsp}d\bar{\psi}\right],
   \label{fe:Ed-action:Eq}
}
where $R(\bMsp)$ is the Ricci scalar corresponding to the conformally transformed metric $w_{PQ}(\bMsp)dv^Pdv^Q = \e^{2[(D-2)A+p\psi]/(n-1)}
\left(q_{\mu\nu}(\Usp) du^{\mu}du^{\nu} + dy^2\right)$.
The Hodge operator on $\bMsp$ space is defined as $\ast_{\bMsp}$ and 
$\tilde{\kappa}$ is given by $\tilde{\kappa}=V^{-1/2}\kappa$ with the 
volume of the internal space $\Zsp$ given by
\Eq{
V\equiv\int_{\Zsp}\ast_{\Zsp}{\bf 1}_{\Zsp}\,.
  \label{fe:volume:Eq}
} 
In obtaining Eq.~(\ref{fe:Ed-action:Eq}), we have dropped 
the surface terms coming from $\lap_{\bMsp}A$, $\lap_{\bMsp}\psi$, 
where $\lap_{\bMsp}$ is the Laplace operator constructed 
from the metric $w_{PQ}(\bMsp)$. 
The potential $V\left(\bar{A},~\bar{\psi}\right)$ is given by
\bea
V\left(\bar{A}, \bar{\psi}\right)&=&U(\bar{A})W(\bar{\psi})~,\label{vuw}
\eea
where
\Eqrsubl{fe:p:Eq}{
U(\bar{A}) &=&\exp\left[-\frac{2(D-2)\bar{A}}{(n-1)\sqrt{c_1}}\right],
\label{ua}\\
W(\bar{\psi})&=&
2\Lambda\exp\left\{-\frac{2p\bar{\psi}}{(n-1)\sqrt{c_3}}\right\}
+\frac{f^2}{2}
\exp\left\{-\frac{2np\bar{\psi}}{(n-1)\sqrt{c_3}}\right\}\nn\\
&&-p\lambda\exp\left\{-\frac{2(D-2)\bar{\psi}}{(n-1)\sqrt{c_3}}\right\}.
     \label{fe:potential:Eq}
}
The fields $\bar{A}$, $\bar{\psi}$ have been rescaled according to 
\Eq{
\bar{A}=\sqrt{c_1}A\,,~~~~~~ 
\bar{\psi}=\sqrt{c_3}\psi\,,
  \label{fe:c:Eq}
}
with the constants $c_i~(i=1, 2, 3)$ defined by 
\Eqrsubl{fe:c13:Eq}{
c_1&=&2\left[\frac{n}{n-1}(D-2)-2(D-1)\right](D-2)
\label{fe:c1:Eq}
\nn\\&&
+2\left[n-1+\frac{2}{\alpha^2}(p-1)\right](p-1)+2p(D-1)\,,
\\
c_2&=&\frac{4(D-2)p}{n-1}\,,
\label{fe:c2:Eq}
\\
c_3&=&2p\left(\frac{n-1}{p}+1\right)\,.
\label{fe:c3:Eq}
}
The absence of a stabilization mechanism for the 
modulus associated to the warp factor 
is clear from the form of the potential in \eqref{vuw}. 
In the $\bar A$-direction the potential decays exponentially causing the 
modulus $\bar A$ to suffer from a runaway behavior and the warped direction 
to expand forever. In the present set-up, classically, 
the warped direction cannot be stabilized. On the other hand the vacuum 
expectation value of $\bar \psi$ can be fixed by appropriately 
tuning the gauge field (see Fig.~\ref{vpsi}). 
The potential energy at the minimum is equivalent 
to the $(n+1)$-dimensional cosmological constant. 
Since the moduli potential energy eventually
turns out to be positive or negative, the $(n+1)$-dimensional background 
geometry becomes dS${}_{n+1}$ or AdS${}_{n+1}$ spacetime. 
In the following, we will assume that the 
modulus $\bar{\psi}$ is fixed at 
$\bar{\psi}=\bar{\psi}_0$ by tuning the gauge field flux. 
This does not affect the dynamics of the other modulus 
$\bar{A}$, whose stabilization will be considered in the next section.

As far as quantum corrections from the $p$-form field are concerned, these only produce a small in the constant $f$, therefore not affecting the classical stabilization of the modululs $\psi$. As it can be seen from Eqts.~(1.6) such corrections: 1) will not spoil the classical background solution, 2) and will not be able to stabilize the potential for the other modulus $A$. The situation may be different, if one wished to introduce additional moduli by perturbing along other directions the classical background solution. 


\section{Quantum effects from the modulus $A$}
\label{sec:mass}
   
In this section, we will discuss the possibility of stabilizing the 
modulus degree of freedom associated with the warp factor 
in the lower-dimensional effective theory described in Sec. \ref{sec:flux}.
We will adopt the background field method and the path integral 
approach to compute the effective potential at one-loop for the 
moduli-field and deal with the divergences using zeta-function 
regularization.

Let us consider the $(n+1)$-dimensional scalar sector of the action 
(\ref{fe:Ed-action:Eq})
\Eq{
I_{\bar{A}}\left[\bar{\psi}_0,~\bar{A}(u_\mu, y)\right]
=-\frac{1}{2\tilde{\kappa}^2} \int_{\bMsp}
 \left[\frac{1}{2}d\bar{A}\wedge\ast_{\bMsp}d\bar{A}
 +V\left(\bar{A},~\bar{\psi}_0\right)\ast_{\bMsp}{\bf 1}_{\bMsp}\right],
   \label{qc:action:Eq}
}
and expand the field $\bar A$ around its classical vacuum expectation value, 
$\bar{A}_0$ given by (\ref{abg}),
\bea
\bar{A}(u_\mu, y) = \bar{A}_0 + a~,
  \label{qc:A:Eq} 
\eea
with $a$ representing the quantum fluctuation. 
Expanding the action up to second order 
\Eq{
S=\frac{1}{2\tilde{\kappa}^2} \int_{\bMsp}
 \left[\left\{R(\bMsp)-V\left(a,~\bar{A}_0,~\bar{\psi}_0\right)\right\}
 \ast_{\bMsp}{\bf 1}_{\bMsp}-\frac{1}{2}da\wedge\ast_{\bMsp}da\right],
   \label{fe:Ed-action2:Eq}
}
where linear terms in $a(u_\mu, y)$ have disappeared owing to the classical 
equations of motion and the potential 
$V\left(a,~\bar{A}_0,~\bar{\psi}_0\right)$ is given by 
\Eq{
V\left(a,~\bar{A}_0,~\bar{\psi}_0\right)=
U(\bar{A}_0)W(\bar{\psi}_0) \left[ 1 
+\frac{2(D-2)^2}{(n-1)^2c_1}a^2\right]+O(a^3). 
   \label{qc:potential2:Eq}
}
After varying the action with respect to $a$, 
we obtain the field equations for the fluctuation
\Eqr{
\left(\lap_{\bMsp}-M^2_{a}\right)a=0\,,
}
where $M^2_{a}$ is given by 
\Eqr{
M^2_{a}=\frac{2(D-2)^2}{(n-1)^2c_1}
U(\bar{A}_0)W(\bar{\psi}_0)\,.
    \label{fe:mass:Eq}
}
Using path integrals we can express the amplitude as
\Eq{
Z=\int{\cal D}[a]\exp{\left(iI_{\bar{A}}
[\bar{\psi}_0,~\bar{A}]\right)},
       \label{qc:amp:Eq}
}
where ${\cal D}[a]$ is a measure on the functional 
 space of scalar fields $a(u_\mu, y)$, and 
 $I_{\bar{A}}[\bar{\psi}_0,~\bar{A}]$ is given by (\ref{qc:action:Eq}). 
At one-loop, it is sufficient to compute the above path integral with 
the action expanded up to second order around its classical background value,
\Eq{
I_{\bar{A}}
\left[\bar{\psi}_0,~\bar{A}(u_\mu, y)\right]=I_{\rm c}
\left[\bar{\psi}_0,~\bar{A}_0\right]
+I_{\rm q}\left[\bar{\psi}_0,~\bar{A}_0,~a\right]+O\left(a^3\right),
       \label{qc:action2:Eq}      
}
where $\bar{A}(u_\mu, y)$ is given by (\ref{qc:A:Eq}) and linear terms in $a$ 
have disappeared due to the classical equations of motion. 
Using the above expression, the path integral (\ref{qc:amp:Eq}) becomes 
\Eq{
\ln Z=iI_{\rm c}\left[\bar{\psi}_0,~\bar{A}_0\right]
      +\ln\left\{\int{\cal D}\left[a\right]
     \exp{\left(iI_{\rm q}\left[\bar{\psi}_0,~\bar{A}_0,
     ~a\right]\right)}\right\}.
          \label{qc:amp2:Eq}
}
The above integral is ill-defined because the operators in 
Eq.~(\ref{qc:amp2:Eq}) are unbounded from below in the dS${}_{n+1}$ 
spacetime with Lorentz signature. In order to correct this pathology, 
we proceed in the usual way and by performing a Wick rotation re-express 
(\ref{qc:amp2:Eq}) in the Euclidean form,
\Eq{
\ln Z=-I_{\rm cE}[\bar{\psi}_0,~\bar{A}_0]+\ln\left\{\int{\cal D}[a]
       \exp{\left(-I_{\rm qE}[\bar{\psi}_0,~\bar{A}_0,~a]\right)
       }\right\},
          \label{qc:amp3:Eq}
}
where $I_{\rm qE}$ is the Euclidean action expressed by  
\Eq{
I_{\rm qE}[\bar{\psi}_0,~\bar{A}_0,~a]
=\frac{1}{4\tilde{\kappa}^2}\int_{\bMsp} a
 \left[-d\ast_{\bMsp}d
 +M^2_{a}\ast_{\bMsp}{\bf 1}_{\bMsp}\right]a\,.
           \label{qc:action4:Eq}
} 
Here we have integrated by parts over the kinetic term.

The one-loop quantum effective potential $V_{\rm q}$ is defined according 
to the relation  
\Eqr{
\exp\left[-\int_{\bMsp} V_{\rm q}(\bar{\psi}_0,~\bar{A}_0) 
\ast_{\bMsp}{\bf 1}_{\bMsp}\right]
    &=&\int {\cal D}\left[a\right]
    \exp\left(-I_{\rm qE}[\bar{\psi}_0,~\bar{A}_0,~a]\right)\nn\\
    &=&\left[\det\mu^{-2}\left(\lap_{\bMsp}-M^2_{a}\right)
     \right]^{-\frac{1}{2}},
  \label{qc:qp:Eq}
}
where $\lap_{\bMsp}$ denotes the Laplace operator on $(n+1)$-dimensional 
de Sitter spacetime, and $\mu$ is a normalization constant with dimension 
of mass. Defining
\Eq{
\exp\left[-\int_{\bMsp} V_{\rm q}(\bar{\psi}_0,~\bar{A}_0)
\ast_{\bMsp}{\bf 1}_{\bMsp}\right]
   =\exp\left[-\Omega_{\rm vol}V_{\rm q}(\bar{\psi}_0,~\bar{A}_0)
\right],
  \label{qc:qp2:Eq}
}
with $\Omega_{\rm vol}$ being the volume of $(n+1)$-dimensional de Sitter 
spacetime, we obtain the following expression
\Eq{
V_{\rm eff}(\bar{\psi}_0,~\bar{A}_0)=V_0(\bar{\psi}_0,~\bar{A}_0)
+\frac{1}{2\Omega_{\rm vol}}
             \ln\det\left[\mu^{-2}\left(\lap_{(\bMsp)}
             -M^2_{a}\right)\right],
  \label{qc:ep:Eq}
}
where the above functional determinant has to be evaluated on dS${}_{n+1}$.

A natural way to proceed is to use zeta regularization techniques.
Defining the following generalized zeta function
\Eqr{
\zeta_{a}(s)&\equiv&\sum_{\lambda}        
           \left(\lambda+M^2_{a}
          \right)^{-s}, 
  \label{qc:zs:Eq}
}
where $\lambda$ are the eigenvalues of the Laplacian on dS${}_{n+1}$, 
the effective potential (\ref{qc:ep:Eq}) can be expressed as 
\Eq{
V_{\rm eff}(\bar{\psi}_0,~\bar{A}_0)=V_0(\bar{\psi}_0,~\bar{A}_0)
-\frac{1}{2\Omega_{\rm vol}}
             \left[{\zeta_{a}}'(0)
             +2\zeta_{a}(0)\ln(\mu b)\right],      
  \label{qc:ep2:Eq}
}
where $b$ is the radius of a $(n+1)$-dimensional sphere S${}^{n+1}$.
The task is then to find the analytically continued values of the zeta 
function and its derivative, $\zeta_{a}(0)$ and $\zeta_{a}^{~\prime}(0)$.
The one-loop effective potential can be computed in a variety of ways. 
The most advantageous one is to use contour integral techniques,
which will be done in the reminder of this section. 
However, to see the overall feature of the effective potential,
the simplest way would be the `Schwinger-De Witt' approximation,
which will be performed in Appendix (\ref{SDW}).

The $(n+1)$-dimensional de Sitter geometry, dS${}_{n+1}$, is a 
$(n+1)$-dimensional manifold 
with constant curvature and has a unique Euclidean section
S${}^{n+1}$ with a radius $b$. We call the eigenvalues of the 
Laplacian on this spacetime $\lambda(\ell)$ 
and their degeneracy $d(\ell)$. These are explicitly given by 
\cite{Rubin:1984tc}
\Eq{
d(\ell)=\frac{(2\ell+n)(\ell+n-1)!}{n!\,\ell!}\,,~~~~~
\lambda(\ell)=\ell(\ell+n)\,.
\label{23}
}
Using the generalized zeta function Eq.~(\ref{qc:zs:Eq})
which can be explicitly re-expressed as
\Eqr{
\zeta_a(s)&\equiv&\sum^{\infty}_{\ell=0}d(\ell)        
           \left[\frac{\lambda(\ell)}{b^2}+M^2_a
          \right]^{-s}, 
  \label{vo:zeta:Eqs}
}
the effective potential is given by Eq. (\ref{qc:ep2:Eq}).
We will evaluate the analytically continued values of the zeta function 
(\ref{vo:zeta:Eqs}) at $s=0$ referring to the method employed in 
Refs.~\cite{Kikkawa:1984rx, Kikkawa:1984qc}.

\begin{figure*}[ht]
\unitlength=1.1mm
\begin{center}
\begin{picture}(155,40)
    \put(6.4,17){\rotatebox{0}{\tiny{$-1$}}}
   \put(18,17){\rotatebox{0}{\tiny{$+1$}}}
   \put(78,23.5){\rotatebox{0}{\small{$+\imath$}}}
   \put(78.,15.9){\rotatebox{0}{\small{$-\imath$}}}
  \includegraphics[height=4.3cm,angle=0]{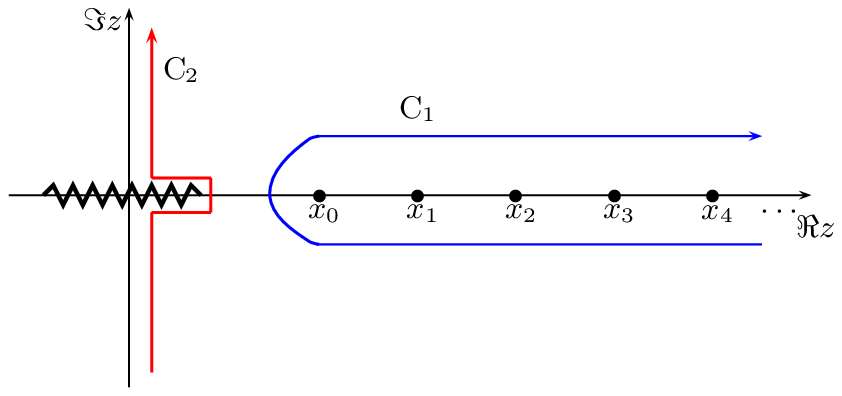}
  \includegraphics[height=4.3cm,angle=0]{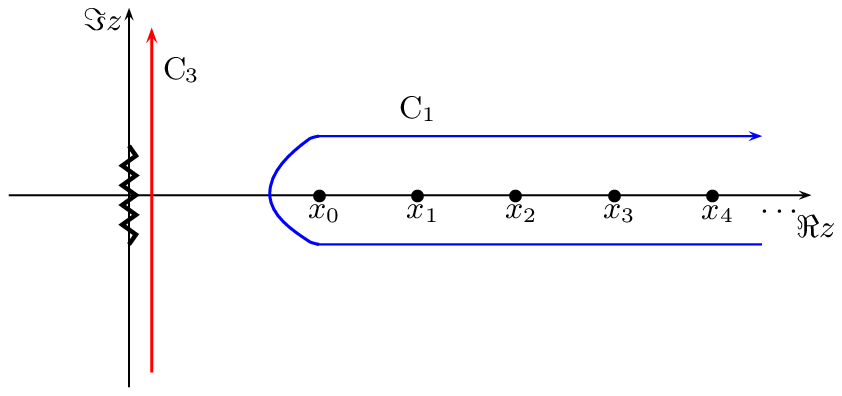}
\end{picture}
\caption{The left panel shows the deformation of the contour ${\rm C}_1$ 
used in (3.22), 
replaced by the contour 
${\rm C}_2$ that avoids the branch points at $z=\pm 1$. 
The right panel shows the deformation of the contour ${\rm C}_1$ used in 
(3.29), replaced by the contour ${\rm C}_3$ running parallel 
to the imaginary axis. The points $x_k$ are defined as:
$x_k=B_N^{-1}\left(N+k+1/2\right)$ for (3.25)
and $x_k=B_{\bar N}^{-1}\left(N+k-1\right)$ for (3.29).
}
  \label{fig:c1}
\end{center}
\end{figure*}

We perform the analytic continuation of the generalized zeta function
to $s= 0$
in the case of $n$ being an odd positive integer,
since the value we are interested in is $n=3$.
Then,
\Eqr{
\zeta_{a}(s)=\sum^{\infty}_{\ell=0}\frac{(2\ell+n)(\ell+n-1)!}{n!\,\ell!}
        \left[\frac{\ell(\ell+n)}{b^2}+M^2_{a}
          \right]^{-s}\,.
       \label{vo:g-zeta:Eqs}
 }
Defining $N=(n+1)/2$ and 
using it as running variable $L=\ell+N$, we rewrite the above expression as  
\Eq{
\zeta_{a}(s)=\sum^{\infty}_{L=N}D_n\left(L-\frac{1}{2}\right)
         \left[\frac{\Lambda_n\left(L-\frac{1}{2}\right)}{b^2}
         +M_a^{2}\right]^{-s},~\label{vo:g-zeta2:Eqs}
}
where we have defined
\Eqrsubl{vo:g-zeta3:Eq}{
D_n\left(L-\frac{1}{2}\right)&=&
             \frac{2L-1}{(2N-1)!}\left[\left(L-\frac{1}{2}\right)^2
             -\left(N-\frac{3}{2}\right)^2\right]\nn\\
             &&\times\cdots \times \left[\left(L-\frac{1}{2}\right)^2
             -\left(\frac{1}{2}\right)^2\right]\,,
        \label{vo:dn:Eq}\\
\Lambda_n\left(L-\frac{1}{2}\right)&=&\left(L-\frac{1}{2}\right)^2
             -\left(N-\frac{1}{2}\right)^2\,.
        \label{vo:ln:Eqs}
}
Using the residue theorem, we can replace the infinite mode sum over $L$ 
by complex integration, obtaining
\Eq{
Z_\pm(s)=-\frac{i}{2}\left(\frac{b}{B_{\rm N}}\right)^{2s}
    B_{\rm N}\int_{{\rm C}_1}dz\tan(B_{\rm N}\pi z)
D_n(B_{\rm N}z)\left(z^2 \mp 1\right)^{-s}\,,
       \label{vo:g-zeta4:Eqs}
 }
where the contour ${\rm C}_1$ in the complex plane is showed 
in Fig.\,\ref{fig:c1}, and $B_{\rm N}^2$ is defined by 
\Eqr{
B_{\rm N}^2&=&\left(N-\frac{1}{2}\right)^2
      -\left(b\,M_a\right)^2\,,
      \label{vo:g-zeta-A:Eqs}
}
(For a positive 
$B_{\rm N}^2$, $-\big(N-\frac{1}{2}\big)<bM_{a}<N-\frac{1}{2}$). 
For clarity, we will consider the two cases separately,
\Eqr{
\zeta_{a}(s)&=&\left\{
\begin{array}{cc}
 Z_+(s)&~{\rm if}~~B_{\rm N}^2>0\,,\\
 Z_-(s)&~~{\rm if}~~B_{\rm N}^2<0\,.
\end{array} \right.
 \label{vo:g-zeta5:Eqs}
   }
Let us consider $Z_+(s)$ first. 
In order to avoid the branch points $z=\pm 1$, 
we may proceed by deforming the contour ${\rm C}_1$ into ${\rm C}_2$ 
as indicated in Fig.~\ref{fig:c1} (left panel), and express $Z_+$ as
\Eqr{
Z_+(s)&=&\left(\frac{b}{B_{\rm N}}\right)^{2s}B_{\rm N}\left[\frac{i}{2}
      \left({\rm e}^{-i\pi s}
     +{\rm e}^{i\pi s}\right)\int^{\infty}_0dx\,D_n(iB_{\rm N}x)
      \left(x^2+1\right)^{-s}\tanh(B_{\rm N}\pi x)\right.\nonumber\\
    & &-\frac{i}{2}\,\e^{i\pi s}\int^1_0
     dx\:\tan \left\{B_{\rm N}\pi(x-i\epsilon)\right\}
     D_n(B_{\rm N}x)\left(1-x^2\right)^{-s}
     \nonumber\\
    & &\left.+\frac{i}{2}\,\e^{-i\pi s}\int^1_0
     dx\:\tan \left\{B_{\rm N}\pi(x+i\epsilon)\right\}
     D_n(B_{\rm N}x)\left(1-x^2\right)^{-s}
      \right]\,,
       \label{vo:z:Eqs}
 }
where $D_n(iB_{\rm N}x)$ defines the following polynomial with coefficients $r_{Nk}$
\Eqr{
D_n(iB_{\rm N}x)&=&i(-1)^{N-1}\frac{2B_{\rm N}x}{(2N-1)!}
        \left[\left(B_{\rm N}x\right)^2
        +\left(N-\frac{3}{2}\right)^2\right]\cdots
        \left[\left(B_{\rm N}x\right)^2
        +\left(\frac{1}{2}\right)^2\right]\nonumber\\
       &\equiv&i(-1)^{N-1}\sum^{N-1}_{k=0}\;r_{Nk}\;
       \left(B_{\rm N}x\right)^{2k+1}\,.
       \label{vo:d2:Eqs}
}
The first term in Eq.\,(\ref{vo:z:Eqs}) comes from the integral along 
the imaginary axis. The second and third terms in Eq.\,(\ref{vo:z:Eqs}) 
are the contributions from the contour along the cut on the real axis. 
Using in the first term of (\ref{vo:z:Eqs}) the following relation
\Eq{
\tanh(B_{\rm N}\pi x)=1-\frac{2}{{\rm e}^{2B_{\rm N}\pi x}+1}\,,
       \label{vo:tanh:Eqs}
}
we arrive at   
\Eqr{
Z_+(s)&=&-\left(\frac{b}{B_{\rm N}}\right)^{2s}\left[\cos(\pi s)
      \frac{1}{\Gamma(s)}(-1)^{N-1}\sum^{N-1}_{k=0}r_{Nk}
      \left(B_{\rm N}\right)^{2k+2}
      \Gamma(k+1)\Gamma(s-k-1)\right.
      \nonumber\\
    & &+i\,B_{\rm N}\,\cos(\pi s)\int^{\infty}_0 dx\,D_n(iB_{\rm N}x)
       \left(x^2+1\right)^{-s}
       \frac{2}{{\rm e}^{2B_{\rm N}\pi x}+1}
      \nonumber\\
    & &\left.+B_{\rm N}\sin(\pi s)\int^1_0 dx D_n(B_{\rm N}x)
        \left(1-x^2\right)^{-s}\tan(B_{\rm N}\pi x)\right]\,. 
         \label{vo:z2:Eqs}
}

Next, we consider the function $Z_-(s)$. 
This time, the branch points in the integrand are on the imaginary axis 
at $z=\pm i$. Therefore we deform the contour
as indicated in Fig.~\ref{fig:c1} (right panel) and obtain
\Eqr{
Z_-(s)&=&i\,\left(\frac{b}{|B_{\rm N}|}\right)^{2s}
     |B_{\rm N}|
\int^{\infty}_0 dx\,D_n(i|B_{\rm N}|x)
\tanh(|B_{\rm N}|\pi x)
     \left(1-x^2\right)^{-s}
      \nonumber\\
    &=&\left(\frac{b}{|B_{\rm N}|}\right)^{2s}
\left[
      (-1)^N\Gamma(-s+1)\sum^{N-1}_{p=0}r_{Np}
      \left(|B_{\rm N}|\right)^{2p+2}
     \left\{\frac{\Gamma(s-p-1)}{\Gamma(-p)}\cos(\pi s)
      \right.\right.\nonumber\\
    & &\left.  
     +\frac{\Gamma(p+1)}{\Gamma(2+p-s)}\right\}
     -2i|B_{\rm N}|
       \left\{\cos(\pi s)\int^{\infty}_{1} dx\,D_n(i|B_{\rm N}|x)
       \frac{\left(x^2-1\right)^{-s}}{{\rm e}^{2|B_{\rm N}|\pi x}+1}
     \right.\nn\\
     & &\left.\left.+\int^{1}_0 dx\,D_n(i|B_{\rm N}|x)
       \frac{\left(1-x^2\right)^{-s}}{{\rm e}^{2|B_{\rm N}|\pi x}+1}
       \right\}\right].
         \label{vo:w:Eqs}
 }
The above expressions, (\ref{vo:z2:Eqs}) and (\ref{vo:w:Eqs}), can be 
easily expanded to get the analytically continued 
values $\zeta_{a}(0)$ and $\zeta'_{a}(0)$.

\begin{figure*}[ht]
\unitlength=1.1mm
\begin{center}
\begin{picture}(125,40)
%
\put(-4.,20){\rotatebox{90}{${V}_{\rm eff}$}}
\put(62,20){\rotatebox{90}{${V}_{\rm eff}$}}
\put(32,-2){${\bar M}_a^2$}
\put(100,-2){${\bar M}_a^2$}
\put(39,14){\tiny{$\mu b=3$}}
\put(39,12){\tiny{$\mu b=4.5$}}
\put(39,10){\tiny{$\mu b=6$}}
\includegraphics[height=4.1cm]{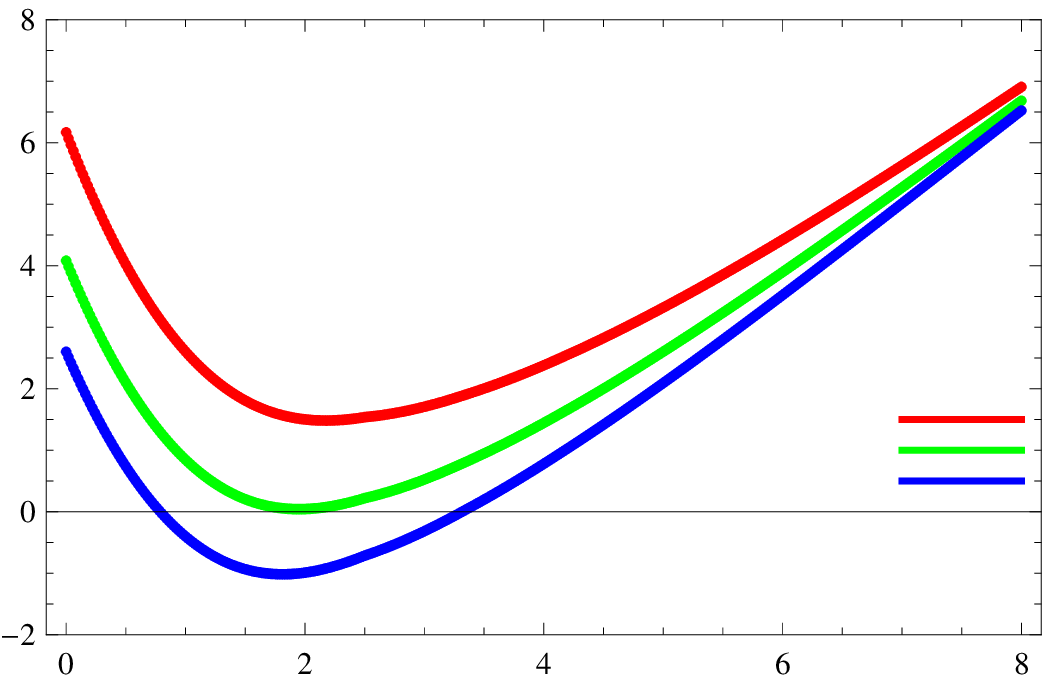}
\hskip 1.0cm
\includegraphics[height=4.1cm]{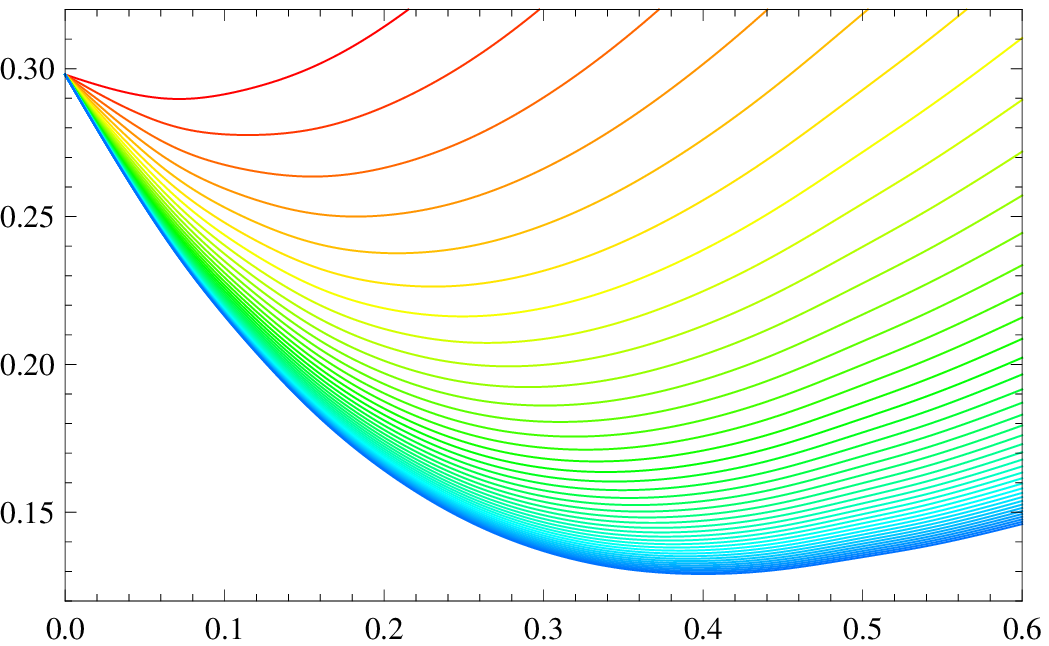}
\end{picture}
\caption{The figures illustrate
the effective potential ${V}_{\rm eff}$ for $\bar{M}_a^2$.
In the left panel, $\mu b=3,4.5,6$ from the top (red), 
while fixing $\alpha=1$ and $b=0.4$.
In the right panel, $\alpha=1.0-5.0$ from the top (red), 
while fixing $b=1.0$ and $\mu b=1.0$.
}
\label{potential_a}
\end{center}
\end{figure*}

\begin{figure}[ht]
\unitlength=1.1mm
\begin{center}
\begin{picture}(70,40)
\put(-6.,20){\rotatebox{90}{${V}_{\rm eff}$}}
\put(65,0){${\bar M}_a^2$}
\includegraphics[height=4.4cm]{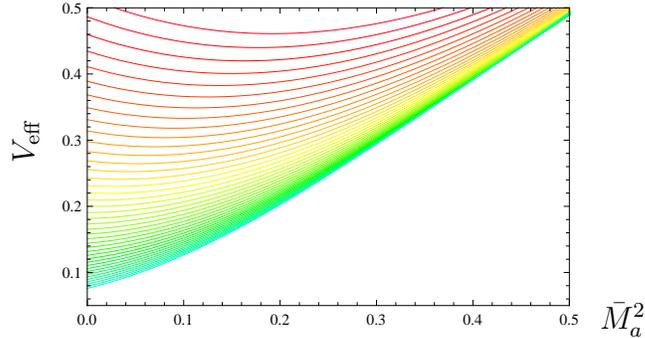}
\hskip 1.0cm
\end{picture}
\caption{The figure illustrates ${V}_{\rm eff}$ for $\bar{M}_a^2$
for changing $b=0.8-1.2$ from the top (red) 
and fixing $\alpha=1.0$ and $\mu b=1.0$.
}
\label{potential_a2}
\end{center} 
\end{figure}

In Figs. \ref{potential_a}-\ref{potential_a2}, for $n=3$,
the behavior of $\bar{V}_{\rm eff}$
is numerically illustrated as a function of ${\bar M}_a^2$,
with three parameters
$\mu$ (or dimensionless $\mu b$), $\alpha$ and $b$.
In the left panel of  Fig. \ref{potential_a},
the effective potential is shown 
for various $\mu b$  
while fixing $b=0.4$ and $\alpha=1.0$,
and in the right panel
it is shown
for various $\alpha$
while fixing  
$b=1.0$ and $\mu b=1.0$.
On the other hand,
in Fig. \ref{potential_a2}, 
it is shown
for various $b$
while fixing $\alpha=1.0$, $\mu b=10$.
For a decreasing $\mu b$ with fixed other parameters
an AdS vacuum is lifted to de Sitter or Minkowski one.
If $\alpha$ is below a critical value for a given set of other parameters,
it is not possible to find a vacuum.
Finally, for an increasing $b$ with fixed other parameters,
the energy density of the de Sitter minimum decreases
but the potential minimum eventually disappears
before it becomes a Minkowski or AdS vacuum.

Similarly, if there is an AdS vacuum,
as $b$ increases, the minima is lifted
but eventually disappears
before it becomes a de Sitter or Minkowski vacuum.
The results of this subsection
are confirmed
by those obtained using the `Schwinger-De Witt' approximation
as described in appendix \ref{SDW}.

In the next section, we will consider the contribution to the potential 
of the quantum fluctuations of the volume 
modulus around the classical minimum $\bar{\psi}=\bar{\psi}_0$
determined by effects of the gauge flux and of the bulk cosmological 
constant.

\section{Quantum contribution to the effective potential from 
the volume-modulus}
\label{sec:mod}

In this section, we consider the case when the size of the internal 
space approaches the Planck length. In this case, 
quantum corrections can no longer be neglected. 
If the size of the internal space is larger than the Planck length,  
quantum effects can be analysed using the conventional loop expansion. 
In the opposite case, the loop expansion breaks down.
Therefore, in the following, we assume that the 
radius of the extra dimensions is larger than the Planck length,
which can provide a natural cut-off scale to the quantum field theory.

Even if stabilized by flux, the volume modulus $\bar\psi$ may still 
contribute to the dynamics of the moduli associated to the warp factor, 
$\bar{A}$ through the coupling of
the quantum fluctuation of $\bar{\psi}$ 
to $\bar{A}$. In this section, 
using the contour integral method,
we will compute the quantum contribution of the modulus $\psi$ 
at one-loop
and discuss whether they can stabilize $\psi$. 
As for the quantum corrections of ${\bar A}$, 
we can expand $\bar{\psi}(u_\mu, y)$ around  
a neighborhood of the local minimum of the potential $W(\bar\psi)$, 
\Eqr{
\bar{\psi}(u_\mu, y)=\bar{\psi}_0+\varphi(u_\mu, y)\,,
  \label{vo:psi:Eq}
}
where $\bar{\psi}_0$ is fixed owing to 
the gauge flux (see Sec. II B). 
The $(n+1)$-dimensional action (\ref{fe:Ed-action:Eq}) expanded up to 
quadratic order in $\varphi(u_\mu, y)$ is 
\Eq{
S=\frac{1}{2\tilde{\kappa}^2} \int_{\bMsp}
 \left[\left\{R(\bMsp)-V\left(\bar{A},~\bar{\psi}_0,~\varphi\right)\right\}
 \ast_{\bMsp}{\bf 1}_{\bMsp}
 -\frac{1}{2}d\varphi\wedge\ast_{\bMsp}d\varphi\right],
   \label{vo:action:Eq}
}
where the potential 
$V\left(\bar{A},~\bar{\psi}_0,~\varphi\right)$ is given by 
\Eq{
V\left(\bar{A},~\bar{\psi}_0,~\varphi\right)=U(\bar{A})
\left[W_0\left(\bar{\psi}_0\right)
+W_2\left(\bar{\psi}_0\right)\varphi^2\right]+O(\varphi^3)\,. 
   \label{vo:potential:Eq}
}
Note that
linear terms disappear owing to the classical equation of motion
and the second term explicitly denotes
the coupling of the quantum fluctuations of the volume modulus $\varphi$
to the warp factor ${\bar A}$.
In the above expression, the functions $
W_i(\bar{\psi}_0)~(i=0,2)$ are 
\Eqrsubl{vo:po2:Eq}{
\hspace{-0.2cm}
W_0\left(\bar{\psi}_0\right)&=&
2\Lambda \varepsilon(p)
+\frac{f^2}{2}
\varepsilon(np)
-p\lambda\,
\varepsilon(D-2)\,,
    \label{vo:po0:Eq}\\
\hspace{-0.2cm}
W_2\left(\bar{\psi}_0\right)&=&\frac{2}{(n-1)^2c_3}\left[
2\Lambda p^2\,
\varepsilon(p)
+\frac{(npf)^2}{2}
\varepsilon(np)
-p\lambda(D-2)^2
\varepsilon(D-2)
\right],
    \label{vo:po02:Eq}
}
where we have defined
\bea
\varepsilon(x)=\e^{-\frac{2x\bar{\psi}_0}{(n-1)\sqrt{c_3}}}~.
\eea
Varying the action with respect to $\varphi$ gives
\Eqr{
\left(\lap_{\bMsp}-M^2_{\varphi}\right)\varphi=0\,,
}
with $M^2_{\varphi}$ is expressed by the relation 
\Eqr{
M^2_{\varphi}=
U(\bar{A})
\,
W_2\left(\bar{\psi}_0\right)\,.
    \label{vo:mass:Eq}
}

The calculation of the one-loop effective potential can be carried out 
using path integrals, and similar steps to those used in the previous 
section allow us to obtain
\Eqr{
V_{\rm eff}(\bar{A},~\bar{\psi}_0)&=&
U(\bar{A})W_0(\bar{\psi}_0)
+V_{\rm q}(\bar{A},~\bar{\psi}_0)\nn\\
&=&U({\bar A})
W_0(\bar{\psi}_0)+\frac{1}{2\Omega_{\rm vol}}
             \ln\det\left[\mu^{-2}\left(\lap_{\bMsp}
             -M^2_{\varphi}\right)\right],
  \label{vo:ep:Eq}
}
where $\lap_{\bMsp}$ denotes the Laplace operator on 
$(n+1)$-dimensional de Sitter spacetime, and $\mu$ 
is a normalization constant with dimension of mass.
The $(n+1)$-dimensional de Sitter geometry, dS${}_{n+1}$, 
is a $(n+1)$--dimensional manifold with constant curvature 
and has a unique Euclidean section
S${}^{n+1}$ with a radius $b$. 
In the following 
we will evaluate the potential 
by analytically continuing
the generalized zeta function
\Eqr{
\zeta_{\varphi}(s)&\equiv&\sum^{\infty}_{\ell=0}d(\ell)        
           \left[\frac{\lambda(\ell)}{b^2}+M^2_{\varphi}
          \right]^{-s}, 
  \label{vo:zeta:Eq2}
}
to $s\to 0$.
The effective potential $V_{\rm eff}(\bar{A},~\bar{\psi}_0)$
is then expressed as 
\Eq{
V_{\rm eff}(\bar{A},~\bar{\psi}_0)=U({\bar A})W_0(\bar{\psi}_0)
             -\frac{1}{2\Omega_{\rm vol}}
             \left[{\zeta_{\varphi}}'(0)
             +2\zeta_{\varphi}(0)\ln(\mu b)\right].      
  \label{vo:ep2:Eq}
}
We will refer to the method employed in Refs.~\cite{Kikkawa:1984rx, 
Kikkawa:1984qc}. 
The contribution of the quantum correction played an important role 
to the effective potential. 
We find that the quantum effective potential 
has a terms proportional to $M_{\varphi}$. 
The procedure is the same
as that employed in Sec. III. C,
except for the replacement of $M_a^2\to M_{\varphi}^2$.

As before, here we will focus on 
the case of $n$ odd and integer.
Using the residue theorem, 
and defining
\Eqr{
B_{\rm N}^2&=&\left(N-\frac{1}{2}\right)^2
      -\left(b\,M_{\varphi}\right)^2\,,
      \label{vo:g-zeta-A:Eq}
}
($B_{\rm N}^2$ is positive for 
$-\big(N-\frac{1}{2}\big)< bM_{\varphi}<N-\frac{1}{2}$),
we will consider the two cases separately,
\Eqr{
\zeta_{\varphi}(s)&=&\left\{
\begin{array}{cc}
 Z_+(s)&~{\rm if}~~B_{\rm N}^2>0\,,\\
 Z_-(s)&~~{\rm if}~~B_{\rm N}^2<0\,.
\end{array} \right.
 \label{vo:g-zeta5:Eq}
   }
Following the same procedure as that in Sec. III C,
we can finally 
reduce $Z_{+} (s)$ and $Z_-(s)$
to the same forms as Eqs. (\ref{vo:z2:Eqs}) and (\ref{vo:w:Eqs}),
respectively,
with the replacement of 
the definition of $B_N^2$ as Eq. (\ref{vo:g-zeta-A:Eq}).

\begin{figure*}[ht]
\unitlength=1.1mm
\begin{center}
\begin{picture}(120,45)
%
%
   \put(-5.,20){\rotatebox{90}{$V_{\rm eff}\left(\bar A\right)$}}
   \put(63,20){\rotatebox{90}{$V_{\rm eff}\left(\bar A\right)$}}
   \put(29,-2){$\bar A$}
   \put(102,-2){$\bar A$}
   \put(46,30){\tiny{$W(\bar\psi_0)$}}
   \put(45,14){\tiny{$\bar\psi_0$}}
  \includegraphics[height=3.9cm]{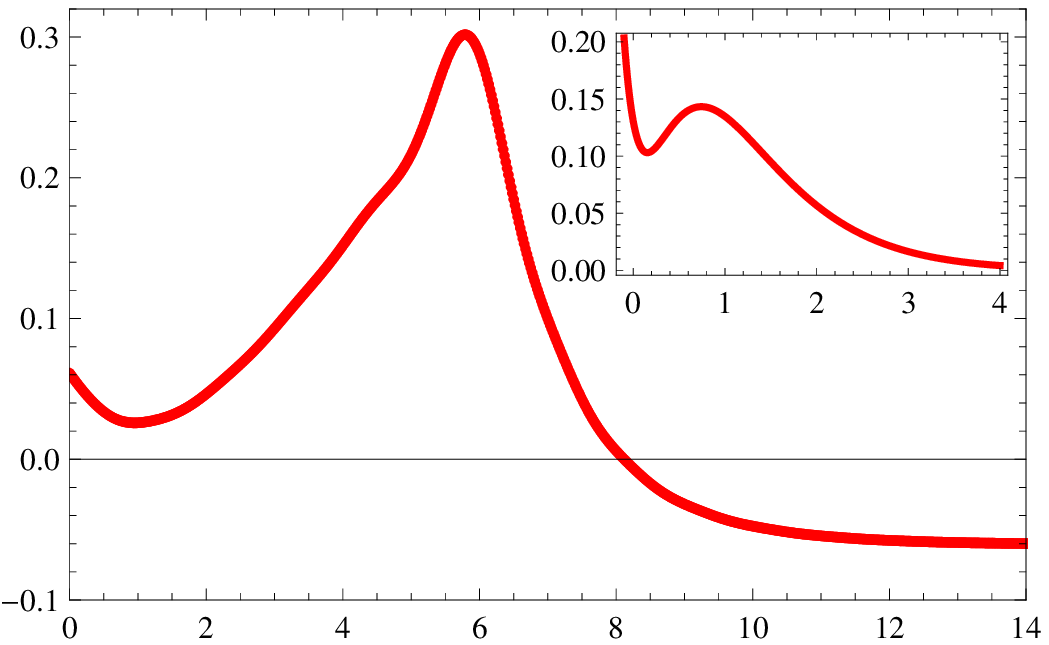}
  \hskip 1.0cm

  \includegraphics[height=3.9cm]{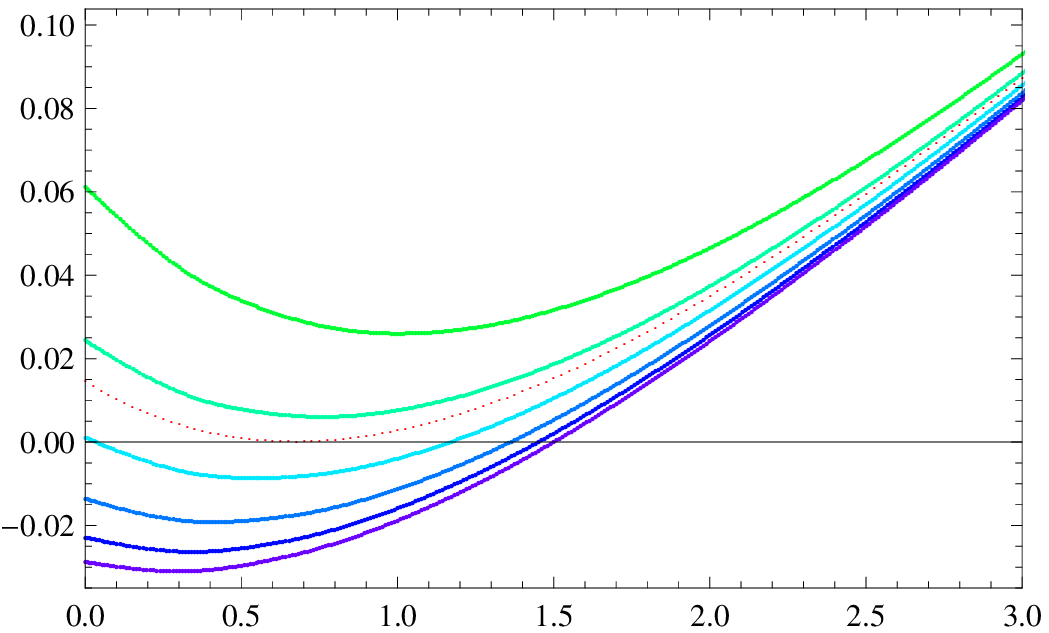}
\end{picture}
  \caption{The figure in the left panel shows a typical 
configuration realizing a de Sitter minima after quantum stabilization. 
The small superposed figure represents the classical potential 
$W(\bar\psi_0)$ for the volume modulus $\bar\psi$ after flux stabilization. 
In the left panel we have set: $\Lambda=1.2$, $\lambda=0.547$, $f=1.6$ 
and $\alpha=1$. The right hand panel shows how the minima of the potential 
depends on the value of $W(\bar \psi_0)$. 
The top green curve corresponds to $W(\bar \psi_0)=10^{-1}$ realizing a 
de Sitter vacuum, while the bottom purple curve corresponds to 
$W(\bar \psi_0)=10^{-2}$ realizing an anti de Sitter vacuum.
The red dotted line corresponds to $W(\bar \psi_0)=0.053$ and realized 
a Minkowski vacuum. }
  \label{potential_1}
\end{center}
\end{figure*}

\begin{figure*}[ht]
\unitlength=1.1mm
\begin{center}
\begin{picture}(120,35)
%
   \put(-7.,16){\rotatebox{90}{$V_{\rm eff}\left(\bar A\right)$}}
   \put(60,16){\rotatebox{90}{$V_{\rm eff}\left(\bar A\right)$}}
   \put(32,-2){$\bar A$}
   \put(103,-2){$\bar A$}
   \put(106,32.8){\tiny{$\alpha=0.1$}}
   \put(106,30.3){\tiny{$\alpha=10$}}
   \put(106,27.8){\tiny{$\alpha=100$}}
   \put(13.5,32.4){\tiny{$\alpha=0.6$}}
   \put(13.5,30.4){\tiny{$\alpha=2.$}}
   \put(13.5,28.4){\tiny{$\alpha=10$}}
  \includegraphics[height=3.9cm]{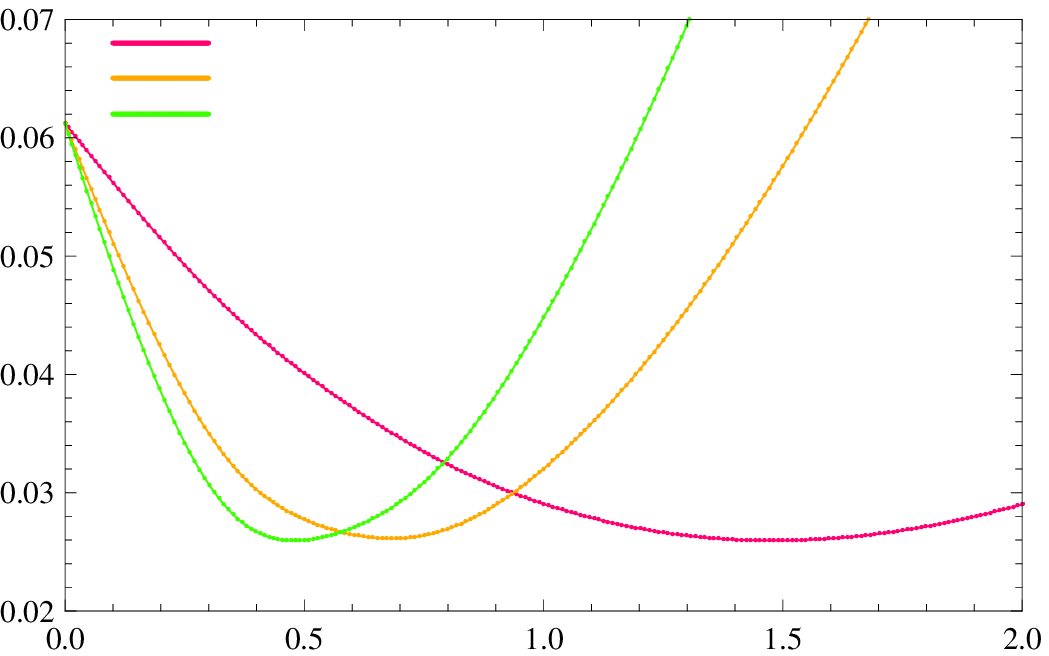}
  \hskip 1.0cm
  \includegraphics[height=3.9cm]{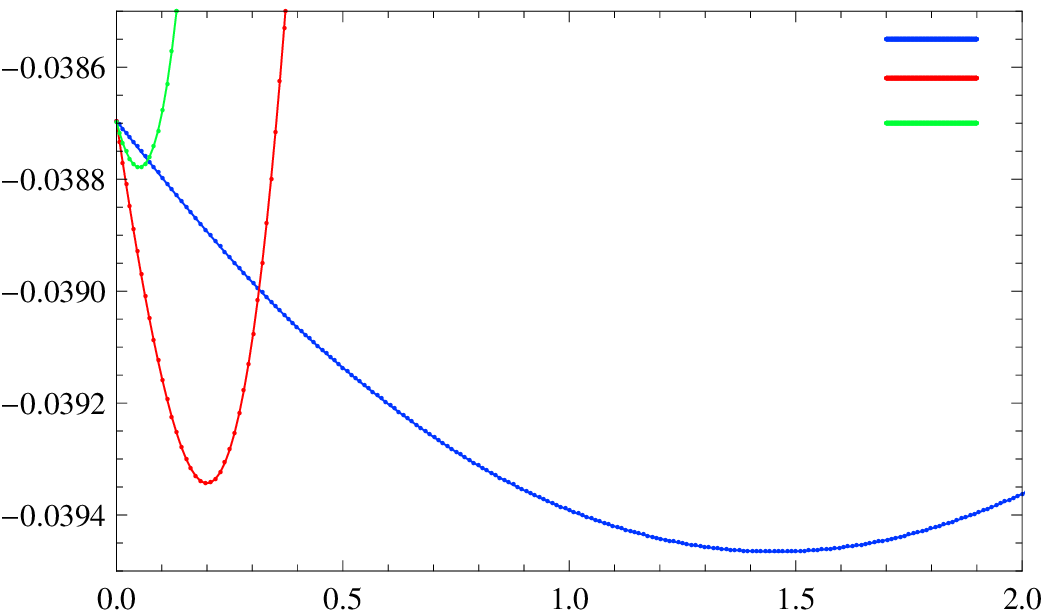}
\end{picture}
  \caption{The figure illustrates the dependence of the potential on 
the parameter $\alpha$. In the left panel we have set the parameters 
$\Lambda$, $\lambda$ and $f$ as in the previous figure in order to obtain, 
after flux stabilization, a positive minima, $W(\bar\psi_0)=10^{-1}$. 
In the right panel we have reduced the flux to obtain 
$W(\bar\psi_0)=10^{-4}$. In the first case (left panel), 
after quantum stabilization the minima is positive realizing a de sitter 
vacua, while in the right panel the minima is negative giving an anti 
de Sitter vacua. Decreasing the parameter $\alpha$ shifts the minima 
towards larger values, without changing the sign of the potential.}
  \label{potential_2}
\end{center}
\end{figure*}

\begin{figure*}[ht]
 \begin{center}
  \includegraphics[height=4.5cm]{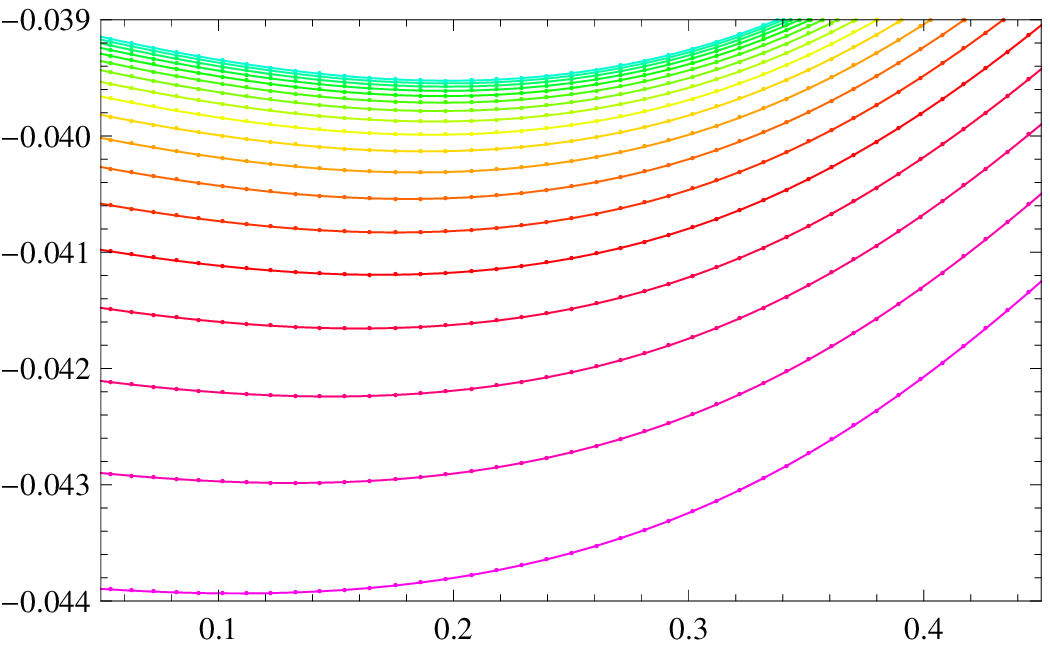}
\put(-230, 70){\rotatebox{90}{$V_{\rm eff}\left(\bar A\right)$}}
\put(5,-2){$\bar A$}
  \caption{In this plot we show how the minima of the potential 
  in the $\bar A$-direction shifts when the flux stabilizes the volume 
modulus to a negative value, generating an AdS vacua. 
  The top curve refer to $W(\bar\psi_0)=-10^{-4}$ while the bottom 
  curve refers to $W(\bar\psi_0)=-5\times 10^{-3}$. One may notice 
  that for negative and decreasing values of $W(\bar\psi_0)$ the 
  minima of the potential accumulates around $V_{\rm min}\sim 0.0395$ for 
  the present choice of parameters. For values of $W(\bar \psi_0)$ 
  below $5\times 10^{-3}$ the minima disappears.
  }
  \label{potential_3}
 \end{center}
\end{figure*}

The above expressions for the zeta functions can be directly 
used to obtain the one-loop effective potential. 
While explicit expressions can be obtained from formulae (\ref{vo:z2:Eqs}) 
and (\ref{vo:w:Eqs}), here we follow a more 
expedite approach based on numerical approximation. 
Results are shown for the case of $n$ odd that we set $n=3$ and 
$B_N^2$ positive. 
Figs.~\ref{potential_1}-\ref{potential_3} illustrate
the effect of the one-loop corrections from quantum fluctuations of 
the volume modulus $\psi$ after flux stabilization. 
Depending on the value that the potential $W(\bar\psi)$ attains at the 
minima, various possibilities 
can be realized. Fig.~\ref{potential_1} shows a typical configuration 
that realizes a de Sitter minima, for positive $W(\bar\psi_0)=10^{-1}$. 
The right panel of Fig.~\ref{potential_1} shows how the potential depends 
on the value of $W(\bar\psi_0)$ illustrating how, for increasing values of 
$W(\bar\psi_0)$ the vacuum can be lifted from AdS to Minkowski or de Sitter. 
The quantum correction basically lifts the potential up 
without changing the shape too much around the minimum, which allows to 
uplifts the AdS minimum and make it a metastable de Sitter ground state.  
Fig.~\ref{potential_2} depicts the dependence of the potential 
on the parameter $\alpha$, showing that a decrease in $\alpha$ tends to 
shift the minima towards larger values. 
Finally, for $W(\bar\psi_0)$ negative, again an AdS vacua is realized and 
increasing $W(\bar\psi_0)$ does not change the sign of the minima of 
the effective potential as long as $W(\bar\psi_0)$ remains negative 
(see Fig.~\ref{potential_3}) and the potential tends to accumulate on 
the upper curve.
For values of $0>W(\bar\psi_0)$ below a certain critical value, it is not 
possible to achieve any minima when quantum effects are included.

The classical potential of $\bar{A}$ forces to decompactify 
the extra dimension while the combinations of matter and quantum correction 
produce a local minimum of the effective potential. Hence, the scale of the 
internal space $\Zsp$ is stabilized by balancing the 1-loop correction, 
gauge field strength wrapped around the internal space and 
the curvature term of the internal space with the cosmological constant.
If we can have a negative potential minimum for a choice of the 
parameters, a dS${}_{n+1}$ spacetime
evolves into a AdS${}_{n+1}$ when
the modulus settles down to the potential minimum.

\section{Discussions}
  \label{sec:Discussions}
In this paper, we have tackled the issue of the moduli stabilization 
in a class of higher dimensional models with two moduli. 
One ($\psi$) is related to the volume of the internal space, 
while the other ($A$) is related to the warped direction. 
These models provide interesting cosmological toy-models owing to the 
fact that it is possible to realize explicit exact de Sitter solutions.

In previous work (see Ref.~\cite{Minamitsuji:2011gp}), the 
lower-dimensional effective theory has been derived, 
with the warped direction regarded as an external one
and the warp factor as a 
modulus. Unfortunately, the lower-dimensional effective theory derived
in Ref.~\cite{Minamitsuji:2011gp}
was problematic due to the runaway behavior of the potential. 
To address this problem here we have discussed a consistent mechanism of stabilization for the warp factor.

The example we have considered is 
simple enough, 
in the sense that only two moduli 
are included in the analysis. 
While the volume modulus can be fixed by 
appropriately tuning the gauge flux, the same mechanism cannot work for the 
modulus associated to the warp factor.
Therefore, in the present paper,
we have discussed whether quantum fluctuation from both moduli 
can lead to full stabilization. 
We have discussed this by using the background field method, 
path-integrals and zeta-function regularization, and showed that, 
quantum effects from 
both moduli may provide an efficient solution to the 
stabilization problem in the present model. 
In the presence of the 1-loop correction, 
the classical contributions from 
curvature and flux compete with quantum effects
leading to a local minimum and showed that by tuning $\alpha$ and 
$\mu$, one can perturb the AdS vacua to produce dS vacua. 
The vacua will clearly only be metastable, since all of the sources of 
energy we have introduced vanish or become negative as 
$\bar{A}\rightarrow\infty$.

\section*{Acknowledgments}
AF acknowledges the support of the Funda\c{c}\~{a}o p\^{a}ra a  
Ci\^{e}ncia e a Tecnologia of Portugal and of the European Union 
Seventh Framework Programme (grant agreement  PCOFUND-GA-2009-246542).  
MM is supported by the Funda\c{c}\~{a}o p\^{a}ra a  Ci\^{e}ncia e a 
Tecnologia of Portugal (SFRH/BPD/88299/2012) and by a Grant-in-Aid for Young 
Scientists (B) of JSPS Research, under Contract No. 24740162.

\appendix

\section{`Schwinger-De Witt' approximation}
\label{SDW}
Here, we will provide a simpler way to compute the one-loop 
effective potential (\ref{qc:ep2:Eq}) directly 
using the Schwinger-De Witt expansion for the heat-kernel. 
This approach is valid in the region of parameter space 
for which the value of $M_a$ is large enough.

Using the Mellin transform, the zeta function can be expressed as
\Eq{
\zeta_{a}=\frac{1}{\Gamma(s)}\int^{\infty}_0dt 
t^{s-1}\e^{-(M^2_{a}+H^2)t}\Theta(t)\,,
   \label{qc:zeta:Eq}
}
where $H:=b^{-1}$ is the Hubble scale of the de Sitter space and 
the function $\Theta(t)$ is the heat-kernel defined as
\Eq{
\Theta(t)=\sum_{\lambda}\e^{-(\lambda -H^2) t}\,.
}
If the value of the mass $M_a$ is large enough, then the exponential 
in the integral above suppresses the contribution coming from the 
large-$t$ part of the integration range, and a direct use of the 
small-$t$ expansion is possible. This procedure is analogous to the 
high temperature expansion of the effective action. After rescaling 
the integral (\ref{qc:zeta:Eq}) by $t\rightarrow H^{-2} \bar{t}$, 
it is straightforward to 
realize that the exponential suppression becomes substantial when 
$M_a^2 H^{-2}$ becomes large enough. Using (\ref{fe:mass:Eq}), it is 
straightforward to see that choosing $\alpha \sim O(1)$ and tuning the 
gauge flux in such a way to obtain $W(\bar\psi_0)\sim O(1)$, a small 
hierarchy between the Hubble parameter $H$ and the Planck mass 
($H\sim 10^{-1} M_{\rm Pl}$) is sufficient to generate enough exponential 
suppression.

In this region we may approximate the integrand in (\ref{qc:zeta:Eq})
by using the Schwinger-De Witt expansion for $\Theta(t)$   
\Eq{
\Theta(t)=\frac{1}{(4\pi t)^{(n+1)/2}}\sum_k\tilde\theta_k t^k\,,
}
where the coefficients $\tilde\theta_k$ are the heat-kernel 
coefficients \cite{Kirsten,Vassilevich:2003xt}. Explicit form for the 
coefficients can be found with little work and for the present case of 
de Sitter space with $\xi_c=\frac{3}{16}$ and $n=3$, these are
\Eq{
{\tilde \theta}_0
=\Omega_{\rm vol},~~
{\tilde \theta}_1
=-\Omega_{\rm vol}
\left(\frac{1}{4}H^2\right),
~~
{\tilde \theta}_2
=-\Omega_{\rm vol}
\left(\frac{17}{480}H^4\right),~~
{\tilde \theta}_3
=-\Omega_{\rm vol}
\left(\frac{457}{40320}H^6\right),
}
where ${\Omega}_{\rm vol}$ is defined by 
\Eq{
{\Omega}_{\rm vol}
=\int d^{n+1}x \sqrt{g}\,.
}
A direct computation gives for the one-loop effective potential for 
$D=10$ and $n=3$ the following expression
\Eqr{
&&\hspace{-1.1cm}
\bar{V}_{\rm eff}(M_{a})= V_0+V_{\rm q}(M_{a})
\nonumber\\
&&~~~= \frac{8\alpha^2+25}{32\alpha^2}
   \bar{M}_a^2
-\frac{1}{32\pi^2} 
\left[\frac{3}{4} \left(\bar{M}_a^2+2\bar{H}^2\right)^2
\right.\nonumber\\
&&\left.~~~~~+\left\{
 \frac{1}{2}\left(\bar{M}_a^2+2\bar{H}^2\right)^2
-\frac{1}{15}\bar{H}^4\right\}
\ln\left(\frac{\bar{\mu}^2}{\bar{M}_a^2+2\bar{H}^2}\right)
-\frac{8}{315}\frac{\bar{H}^6}{\bar{M}_a^2+2\bar{H}^2}
\right]\,,
}
where we have rescaled the various quantities according to
\Eqr{
\bar{V}_{\rm eff}&=&V_{\rm eff}\,\tilde{\kappa}^4\,,~~~~~
\bar{H} = H\tilde{\kappa}\,,\nn\\
\bar{{M}_{a}}&=&M_{a}\tilde{\kappa}\,,~~~~~
\bar{\mu} = \mu\tilde{\kappa}\,.\nn
   \label{qc:p2:Eq}
}
Eventual non-vanishing minima of the potential determine the mass 
of the field $a$:
\Eqr{
0&=&\Delta \bar{V}_{\rm eff}(\alpha,\bar{\mu},\bar{H},\bar{M}_a),\\
0&=&\frac{\partial}{\partial \bar{M}_a^2}
  \Delta \bar{V}_{\rm eff}(\alpha,\bar{\mu},\bar{H},\bar{M}_a),
}
where we have normalized the potential according to
\Eqr{
\Delta \bar{V}_{\rm eff}(\alpha,\bar{\mu},\bar{H},\bar{M}_a)
&:=&
\bar{V}_{\rm eff} - 3{\bar H}^2.
}
For a given set of $(\alpha,\bar{\mu},\bar{H})$, 
the solution for $\bar{M}_a$ leads to  
\Eq{
V_{\rm eff}\simeq 3\tilde{\kappa}^{-4} \bar{H}^2\,.
}
Hence, for $H$ to be ${\bar H}\ll 1$ and 
$(\bar{M}_a/\bar{H})\gg 1$, then if
one keeps ${\bar \mu}=O(1)$ fixed, 
the energy density at the minimum is much smaller 
than the Planck scale, which implies that
the stabilization due to the quantum corrections
is working consistently.
In case of $D=10$ and $n=3$,
the classical potential approaches a constant from above as $\alpha$ 
increases. 
For $\alpha$ tuned to be small but non-zero, 
the quantum correction no longer contribute to the effective potential. 
For modest values of $\alpha$, 
we will find numerically that 
there is a solution ${\bar M}_a \simeq 12$
for $\bar{\mu}\gtrsim 10$.

Approximate expressions for the minima of the potential can be found 
at leading order by expanding for $\bar{M}_a\gg \bar{H}$. 
In this regime the minima is determined by
\Eq{
\bar{M}_a^2
\left[1
    +\ln \left(\frac{{\bar \mu}^2}{{\bar M}_a^2
}\right)
\right]=\frac{\pi^2}{\alpha^2}\left(25+8\alpha^2\right).
\label{minimum}
}
Assuming the renormalization scale to be of the same order as the mass,
$\bar\mu \sim M_a$, 
we find 
\Eq{
\bar{M}_a^2
\simeq 
{\bar \mu}^2
\left[1+
\sqrt{1-\frac{\pi^2}{\alpha^2 {\bar\mu}^2}(25+8\alpha^2)}
\right].
\label{app_min}
}
Higher order corrections do not change the qualitative features of 
the above result.

The value that the potential attains at the minima depends on the 
choice of the renormalization scale. Minimizing $\bar{V}_{\rm min}$
as a function of $\bar{\mu}$ allows to find a Minkowski vacua 
($V_{\rm min}=0$) for 
\Eq{
{\bar \mu}_{\rm crit}^2\simeq 
\frac{4\pi^2}{3\alpha^2} (25+8\alpha^2)
=\frac{4}{3}
\left[
8\pi^2+\left(\frac{5\pi}{\alpha}\right)^2
\right]>{\bar\mu}_{\rm min}^2\,,
   \label{qc:min:Eq}
}
where $\mu_{\rm min}$ is the minimum 
value of the renormalization scale for which a minima 
with positive vacuum energy exists.
An AdS minimum ($V_{\rm min}<0$) is found for values of $\bar\mu$ 
in the range $\bar{\mu}>\bar{\mu}_{\rm crit}$, 
while a de Sitter minimum ($V_{\rm min}>0$) is obtained for values 
of $\bar\mu$ lying in the range
$\bar{\mu}_{\rm min}<\bar{\mu}<\bar{\mu}_{\rm crit}$ and the expansion 
rate is given by 
\Eq{
3\bar{H}^2\simeq
-\frac{{\bar \mu}^4}{64\pi^2}
\left[
1-\frac{3\pi^2}{2\alpha^2{\bar\mu}^2}
(25+8\alpha^2)+\left\{1-\frac{\pi^2}{\alpha^2{\bar\mu}^2}
(25+8\alpha^2)
\right\}^{\frac{3}{2}}
\right].
}
The above arguments, although apply in a specific region of the 
parameter space of the model ($\bar\mu$ and $\alpha$) 
suggest that $a$ can be stabilized by quantum effects.

\begin{figure*}[ht]
\unitlength=1.1mm
\begin{center}
\begin{picture}(122,35)
%
\put(-6.,20){\rotatebox{90}{${V}_{\rm eff}$}}
\put(65,20){\rotatebox{90}{${V}_{\rm eff}$}}
\put(26,-2){${\bar M}_a^2$}
\put(97,-2){${\bar M}_a^2$}
   \put(9,33.5){\tiny{Minkowski minima ($T\sim0$)}}
   \put(9,31){\tiny{dS minima}}
   \put(9,28.5){\tiny{High temperature phase}}
   \put(80,12){\tiny{dS minima}}
   \put(80,9){\tiny{Minkowski minima}}
   \put(80,5.5){\tiny{AdS minima ($T\sim 0.1$)}}
  \includegraphics[height=4.1cm]{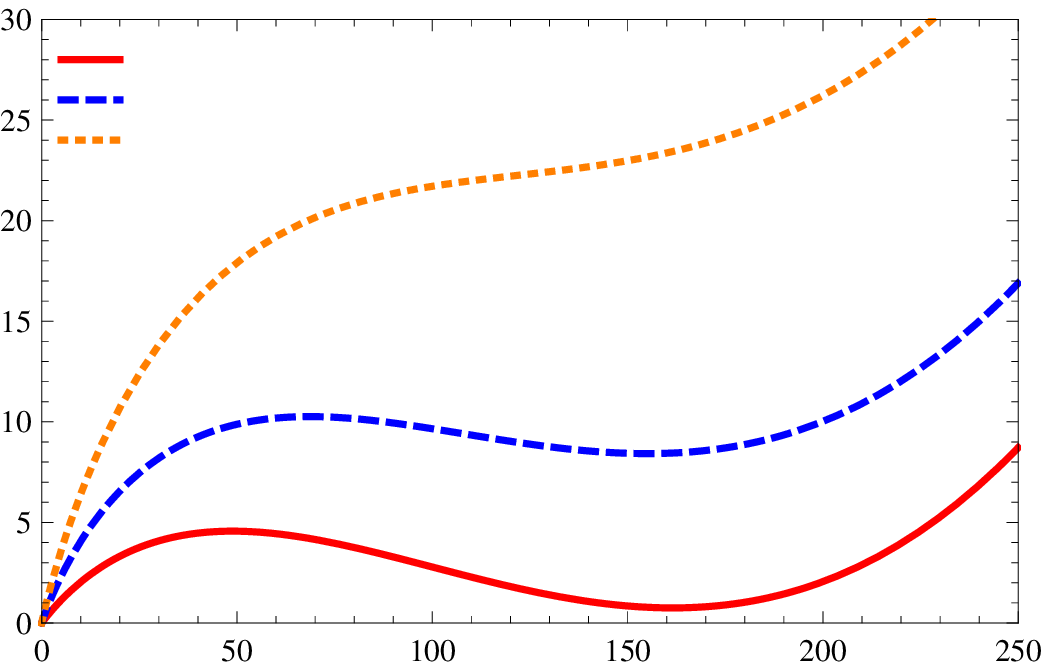}
  \hskip 1.0cm
  \includegraphics[height=4.1cm]{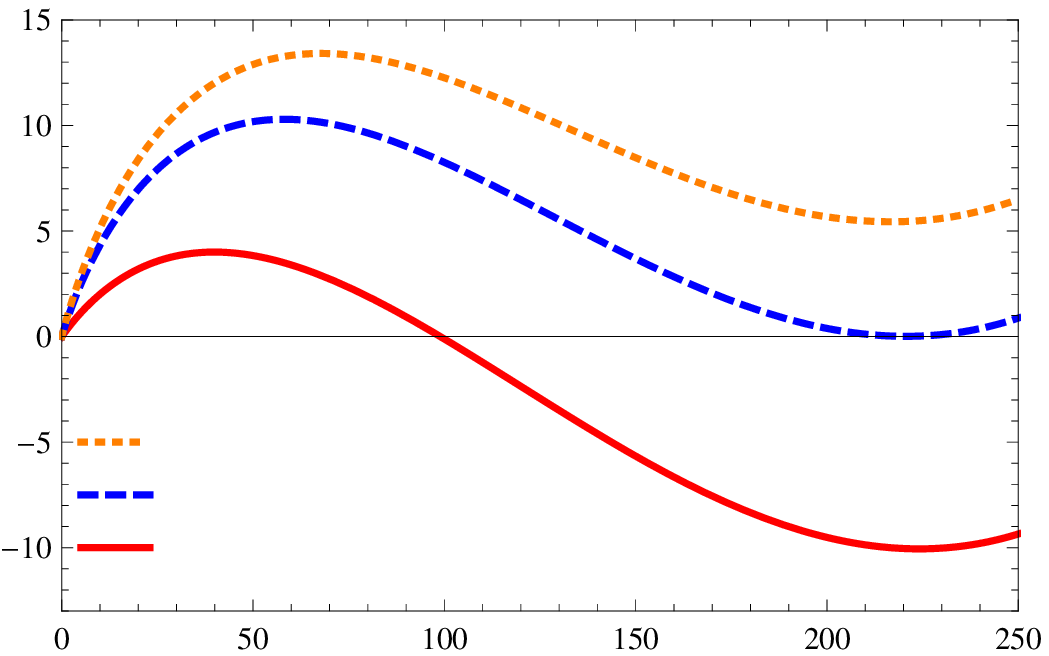}
\end{picture}
  \caption{The left panel illustrates the temperature dependence of 
the potential. The continuous-red curve is tuned to give a vanishing 
vacuum energy at the minima for $T=0$. Increasing the temperature shift 
the minima to a de Sitter vacuum (blue-dashed curve) and further increase 
of the temperature pushes the system into a symmetric high temperature phase 
(yellow-dotted curve). The right panel is illustrates the case in which 
the zero temperature minima is tuned to give a AdS vacuum 
(red-continuous curve), while blue-dotted curve gives a Minkowski minima 
and the yellow-dotted curve the de Sitter minima. 
(Left Panel) The parameters have been set to $\mu=100$ and $\alpha=7$. 
The curves correspond to the following values of the temperature: 
$T=0.1$ (bottom), $T=3.0$ (central), $T=4.0$ (top). (Right Panel) 
The parameters have been set to $\mu=120$, $\alpha=7.$ and the curves 
correspond to the following values of the temperature: $T=0.1$ (bottom), 
$T=3.0$ (central), $T=3.5$ (top).
}
  \label{poten}
\end{center}
\end{figure*}

A more general computation of the one-loop effective potential valid 
in all regions of the parameter space 
was given in Sec.~\ref{sec:mass},
which exhibits a behavior consistent
with the results shown in this Appendix.
The dependence of the potential on the energy scale suggests that 
the inclusion 
of finite temperature effects may lift the minima of the potential. 
Of course, these effects are not directly related to the mechanism 
of stabilization discussed in this paper, 
and clearly a proper inclusion of thermodynamic effects requires care, 
particularly if time dependence is taken into account. 
However, in the approximation that the time evolution of the moduli 
fields is adiabatic, it is possible to give an estimate of these effects 
using the standard Matsubara formalism. 
The argument becomes simpler if the scale of the S${}^{n}$ is approximately 
constant, {\it i.e.} if we assume the adiabatic expansion in the 
direction of S${}^{n}$ after compactification 
of the $(n+1)$-dimensional theory over S${}^{1}$ to 
S${}^{n+1}=$S${}^{1}\times$S${}^{n}$ with $H^{-1}$ being the radius of 
the spatial section S${}^{n}$. The computation of the potential at finite 
temperature carried out in Appendix \ref{fta} gives for $n=3$,
\Eqr{
\bar{V}_{\rm eff}&\simeq& \frac{25+8\alpha^2}{32\alpha^2}\bar{M}_{a}^2
-\frac{1}{32\pi^2}
\left[
\left(\bar{M}_{a}^2+\bar{H}^2\right)^2
\left\{
 \frac{3}{4}
-\frac{1}{2}
\ln \left(\frac{\bar{M}_{a}^2+\bar{H}^2}{{\bar \mu}^2}\right)
\right\}\right.\nn\\
&&\left.
+4\sum_{\ell=1}^{\infty}
\left\{
\frac{
      {\bar\chi}\left(\bar{M}_{a}^2+\bar{H}^2\right)}   
     {\pi^2 \ell^2}
\right\}
K_{-2}
\left(
2\pi \ell\sqrt{\frac{\bar{M}_{a}^2+\bar{H}^2}{\bar{\chi}}}
\right)
\right]\,,
\label{qc:potential:Eq}
}
where $\bar{\chi}=\tilde{\kappa}^{1/2}(2\pi T)^2$
and $T$ is the temperature. 
Details along with high- and low-temperature approximation are obtained 
in Appendix \ref{fta}. Here, we show
the typical behavior of the potential in Fig.~\ref{poten} 
where we have normalized its value by subtracting 
the vacuum energy contribution for $M_a=0$,
which corresponds to the $A\to \infty$ limit.

\section{Finite temperature corrections}
\label{fta}
In this Appendix we present the computations of the finite 
temperature corrections to the effective potential. 
As mentioned in Appendix.~\ref{SDW} we assume that the 
time evolution of the modulus is adiabatic, 
allowing us to use the standard Matsubara formalism \cite{Kapusta:2006pm}. 
In this adiabatic regime, the scale of the S${}^{n}$, $H^{-1}$, 
is assumed to be approximately constant. The same formalism of 
Appendix~\ref{SDW} can be applied and the zeta function becomes
\Eq{
\zeta(s) = \sum_{\lambda,\,N} \left( \lambda + {M}_{a}^2+\chi N^2\right)^{-s}
\,,
}
where $\lambda$ are the eigenvalues of the Laplace operator on the 
$n$-sphere, $N$ is integer, and $\chi$ is defined by $\chi=(2\pi T)^2$.
Mellin-transforming the above expression and 
using the same heat-kernel scheme adopted in 
Appendix~\ref{SDW}, it takes simple steps to arrive at
\Eqr{
\zeta(s)&=&
\frac{1}{(4\pi)^{\frac{n}{2}}\Gamma(s)}
\sqrt{\frac{\pi}{\chi}}
\sum_k{\tilde \theta}_k
\left[\left\{
 \left({M}_{a}^2+H^2\right)^{-k-s+\frac{n+1}{2}}
\Gamma\left(s+k-\frac{n+1}{2}\right)\right\}
\right.\nn\\
&&\left. \!\!+4\sum_{\ell=1}^{\infty}
\left(
\frac{\sqrt{\chi({M}_{a}^2+H^2)}}{\pi\ell}
\right)^{-k-s+\frac{n+1}{2}}
K_{k+s-\frac{n+1}{2}}
\left(2\pi\ell\sqrt{\frac{{M}_{a}^2+H^2}{\chi}}\right)
\right].
   \label{ft:zeta:Eq}
}
The important values $\zeta(0)$ and $\zeta'(0)$ can be computed in 
a straightforward manner from the above expression, leading, for $n=3$, to
\Eqr{
\bar{V}_{\rm eff}&\simeq& \frac{25+8\alpha^2}{32\alpha^2}\bar{M}_{a}^2
-\frac{1}{32\pi^2}
\left[
\left(\bar{M}_{a}^2+\bar{H}^2\right)^2
\left\{
 \frac{3}{4}
-\frac{1}{2}
\ln \left(\frac{\bar{M}_{a}^2+\bar{H}^2}{{\bar \mu}^2}\right)
\right\}\right.\nn\\
&&\left.
+4\sum_{\ell=1}^{\infty}
\left\{
\frac{
      {\bar\chi}\left(\bar{M}_{a}^2+\bar{H}^2\right)}   
     {\pi^2 \ell^2}
\right\}
K_{-2}
\left(
2\pi \ell\sqrt{\frac{\bar{M}_{a}^2+\bar{H}^2}{\bar{\chi}}}
\right)
\right],
\label{ft:potential:Eq}
}
where we have used the definitions (\ref{qc:p2:Eq}), and 
$\bar{\chi}=\tilde{\kappa}^{1/2}\chi$.
The volume factor and the heat kernel coefficients are given by
\Eq{ 
\Omega_{\rm vol}
=\frac{2\pi}{\sqrt{\chi}}\times H^{-3}\Omega_{3}\,,~~~~~
{\tilde\theta}_k=\Omega_{3}{\tilde \gamma}_k H^{2k-3}\,.
}
Here $\Omega_{3}$ is the volume of S${}^3$, and ${\tilde \gamma}_k$ is 
given by 
${\tilde \gamma}_0=1$ and ${\tilde\gamma}_k=0$ ($k=1,2,3$) for an S$^3$.\\ 
Below we obtain the limiting behavior of (\ref{ft:potential:Eq}) assuming 
$\bar{M}_{a}\gg \bar{H}$. At low temperature $\chi\rightarrow 0$, 
the modified Bessel functions decay exponentially as 
$\e^{-2\pi\ell\frac{\bar{M}_{a}}{\sqrt{\bar{\chi}}}}$, and the finite 
temperature corrections become small.
Hence we can recover the result of Appendix~\ref{SDW}
\Eq{
\bar{V}_{\rm eff}
\simeq 
\frac{25+8\alpha^2}{32\alpha^2}\bar{M}_{a}^2
-\frac{1}{32\pi^2}
\bar{M}_{a}^4
\left[
 \frac{3}{4}
-\frac{1}{2}
\ln \left(\frac{\bar{M}_{a}^2}{{\bar \mu}^2}\right)
\right].
  \label{ft:eff:Eq}
}
The effective potential (\ref{ft:eff:Eq}) has a minimum at 
\Eq{
\bar{M}_{a}^2
\simeq 
{\bar \mu}^2
\left[
1+
\sqrt{1-
\frac{1}{{\bar \mu}^2}
\frac{\pi^2(8\alpha^2+25)}{\alpha^2}}
\right],
}
where 
\Eq{
\bar{V}_{\rm min}
\simeq 
-\frac{{\bar \mu}^4}{64\pi^2}
\left[
1-\frac{3}{2{\bar \mu}^2}
\frac{\pi^2(25+8\alpha^2)}{
\alpha^2}
+\left\{
1-\frac{1}{{\bar \mu}^2}
\frac{\pi^2(25+8\alpha^2)}{
\alpha^2}
\right\}^{\frac{3}{2}}
\right].
}
A de Sitter minimum exists in the range 
\Eq{
\frac{\pi^2(25+8\alpha^2)}{\alpha^2}
<{\bar\mu}^2
<\frac{4}{3}
\frac{\pi^2(25+8\alpha^2)}{\alpha^2}.
  \label{ft:ds:Eq}
}
At small temperature, expanding appropriately 
and then summing over $0<\ell \in \mathbb{N}$,
we find
\Eqr{
\bar{V}_{\rm eff}
&\simeq& 
\frac{25+8\alpha^2}{32\alpha^2}\bar{M}_{a}^2
-\frac{1}{32\pi^2}\bar{M}_{a}^4
\left[\frac{3}{4}-\frac{1}{2}
\ln \left(\frac{\bar{M}_{a}^2}{{\bar \mu}^2}\right)
\right].
\label{ft:potential2:Eq}
}
In the high temperature limit, we find
\Eqr{
\bar{U}_{\rm eff}
&\equiv& {\bar V}_{\rm eff}
-\left(-\frac{{\bar\chi}^2}{1440\pi^2}\right)
\nonumber\\
&& \simeq\frac{25+8\alpha^2}{32\alpha^2}\bar{M}_{a}^2
-\frac{1}{32\pi^2}
\left[
\frac{{\bar\chi}^2}{45}
+\left(\bar{M}_{a}^2+\bar{H}^2\right)^2
\left\{ \frac{3}{4}-\frac{1}{2}
\ln \left(\frac{\bar{M}_{a}^2+\bar{H}^2}{{\bar \mu}^2}\right)
\right\}\right.\nn\\
&&\left.+4\sum_{\ell=1}^{\infty}
\left\{
\frac{
      {\bar\chi}\left(\bar{M}_{a}^2+\bar{H}^2\right)}   
     {\pi^2 \ell^2}
\right\}K_{-2}
\left(2\pi \ell\sqrt{\frac{\bar{M}_{a}^2+\bar{H}^2}{\bar{\chi}}}
\right)
\right],
\label{ult3}
}
where we have summed over $\ell$ and proceeded with appropriate 
analytic continuation. The above expression is normalized by subtracting 
the vacuum energy contribution for $\bar{M}_{a}=0$.
In the adiabatic approximation adopted, the effect of increasing the 
temperature is to uplift the minimum of the potential 
without changing its shape around the minimum. 



\end{document}